\documentclass[12pt,twoside,letterpaper]{revtex4}

\usepackage{graphicx}
\usepackage{fullpage}
\usepackage{xspace}
\usepackage{placeins}
\usepackage{multirow}

\newcommand{\mev}{\ensuremath{\mathrm{MeV}}\xspace}
\newcommand{\gev}{\ensuremath{\mathrm{GeV}}\xspace}
\newcommand{\gevc}{\ensuremath{\gev/c}\xspace}

\newcommand{\ns}{\ensuremath{\mathrm{ns}}\xspace}
\newcommand{\sn}[2]{\ensuremath{#1\times10^{#2}}\xspace}
\newcommand{\ds}{\displaystyle}
\newcommand{\lfvkappa}{\ensuremath{\kappa}}
\newcommand{\lfvlambda}{\ensuremath{\Lambda}}

\begin{document}



\vspace*{1in}

\begin{center}

  {\large\bf Feasibility Study for a Next-Generation Mu2e Experiment} \\
  \hfill

K.~Knoepfel$^3$, V.~Pronskikh$^3$, R.~Bernstein$^3$, D.N.~Brown$^5$, R.~Coleman$^3$, C.E.~Dukes$^7$, R.~Ehrlich$^7$, M.J.~Frank$^7$, D.~Glenzinski$^3$, R.C.~Group$^{3,7}$, D.~Hedin$^6$, D.~Hitlin$^2$, M.~Lamm$^3$, J.~Miller$^1$, S.~Miscetti$^4$, N.~Mokhov$^3$, A.~Mukherjee$^3$, V.~Nagaslaev$^3$, Y.~Oksuzian$^7$, T.~Page$^3$, R.E.~Ray$^3$, V.L.~Rusu$^3$, R.~Wagner$^3$, and S.~Werkema$^3$ \\
\hfill

\scriptsize
$^1$ \textit{Boston University, Boston, Massachusetts 02215, USA} \\
$^2$ \textit{California Institute of Technology, Pasadena, California 91125, USA} \\
$^3$ \textit{Fermi National Accelerator Laboratory, Batavia, Illinois 60510, USA} \\
$^4$ \textit{Laboratori Nazionali di Frascati, Istituto Nazionale di Fisica Nucleare, I-00044 Frascati, Italy} \\
$^5$ \textit{Lawrence Berkeley National Laboratory and University of California, Berkeley, California 94720, USA} \\
$^6$ \textit{Northern Illinois University, DeKalb, Illinois 60115, USA} \\
$^7$ \textit{University of Virginia, Charlottesville, Virginia 22906, USA} \\

\normalsize

  \hfill

  Submitted as part of the APS Division of Particles and Fields Community Summer Study \\
  (dated: \today) \\
  \hfill

\end{center}

We explore the feasibility of a next-generation Mu2e experiment that uses Project-X beams to achieve a sensitivity approximately a factor ten better than the currently planned Mu2e facility.

\hfill

\pagebreak

\section{Introduction}
\label{sec:introduction}

The Mu2e experiment, to be hosted at Fermilab, is a flagship component of the U.S.\ Intensity Frontier program~\cite{IF_review} and will search for the charged-lepton-flavor-violating process of coherent muon-to-electron conversion in the presence of a nucleus with a sensitivity of four orders of magnitude beyond current limits~\cite{Mu2eCDR}. 
In the context of the DPF Community Summer Study exercise, we investigate the feasibility of a next-generation Mu2e experiment (Mu2e-II) that uses as much of the currently planned facility as possible and Project-X beams to achieve a sensitivity that is about a factor of ten better than Mu2e.  The factor of ten was chosen because it would allow much of the Mu2e facility to be reused while still providing a significant improvement in sensitivity.  A factor of ten improvement over Mu2e will be interesting regardless of the outcome of Mu2e.  If the Mu2e experiment observes events completely consistent with background expectations, then another factor of 10 improvement in sensitivity opens the door to additional beyond-the-standard-model parameter space.  If Mu2e observes a $3\sigma$ excess, then a Mu2e-II upgrade would be able to definitively resolve the situation.  And if Mu2e discovers charged-lepton-flavor-violating physics, then a Mu2e-II upgrade could explore different stopping targets in an effort to untangle the underlying physics.  By measuring the signal rate using nuclear targets at various $Z$, Mu2e-II would have the unique ability to resolve information about the underlying effective operators that are responsible for the lepton-flavor-violating signal~\cite{Kitano:2002mt,Cirigliano:2009bz}.  

For the improved sensitivity, it is assumed that Project-X-like beams are available.  While we restrict ourselves to considering only modest improvements to the Mu2e sensitivity ($\times 10$), it is important to note that with the full scope of Project-X an additional 1-2 orders of magnitude improvement may be possible.  However, to achieve these additional gains in sensitivity will require significant modifications and upgrades to the currently planned Mu2e facility.  We do not further discuss these potential third-generation experiments here.

In the remainder of this note, we begin with a summary motivating searches for charged-lepton-flavor violation, then briefly describe the plans for the Mu2e experiment, and then discuss the second-generation Mu2e-II experiment.  In Sec.~\ref{sec:simulation} we detail the simulation tools used and the Project-X beam parameters assumed.  We use these simulations to estimate Mu2e-II background yields as described in Sec.~\ref{sec:bgd}.  The upgrades to Mu2e required to realize a further improvement in sensitivity are discussed in Sec.~\ref{sec:upgrades}.  Section~\ref{sec:summary} summarizes our findings and concludes.

\subsection{Charged-lepton-flavor violation}
\label{sec:CLFV}
Since the discovery of the muon, particle physicists have been plagued by questions related to the existence of multiple lepton families:  Why are there multiple generations of leptons?  Is charged-lepton flavor conserved?  If so, why?  These questions have led to a series of experiments aimed at measuring flavor violation in charged-lepton interactions.  To date, no such violation has been experimentally observed, but an ongoing search for these processes plays an important role in addressing some of the fundamental questions of high energy physics~\cite{review}.  

Experiments using muons to search for charged-lepton-flavor violation (CLFV) have been constructed to search for $\mu \rightarrow e\gamma$,  $\mu^+ \rightarrow e^+e^-e^+$, and the coherent $\mu^-N \rightarrow e^-N$ conversion process in nuclei.  Limits on the branching ratios ($Br$) have been set for $\mu \rightarrow e\gamma$ and $\mu^+ \rightarrow e^+e^-e^+$ while the $\mu^-N \rightarrow e^-N$ searches set a limit on $R_{\mu e}$, defined as,
\begin{eqnarray}
  R_{\mu e} &=& \frac
  {\Gamma(\mu^{-}\;  N(A,Z) \to e^{-}\; N(A,Z)}
  {\Gamma(\mu^{-}\; N(A,Z)\to \nu_{\mu}\; N(A,Z-1))},
\end{eqnarray}
where $N(A,Z)$ denotes a nucleus with mass number $A$ and atomic number $Z$.  The numerator corresponds to the rate for the CLFV conversion process and the denominator corresponds to the rate for ordinary muon capture on
the same nucleus.  The present state of CLFV searches using muon decay is:
$Br(\mu^{+}\to{}e\gamma)<5.7\times10^{-13}$ \cite{Adam:2013mnn},
$Br(\mu^{+}\to{}e^{+}e^{-}e^{+})<1\times10^{-12}$
\cite{Bellgardt:1987du}, $R_{\mu{}e}<7\times10^{-13}$
\cite{Bertl:2006up}.  There is potential to improve each of these searches significantly over the next decade. The goal of the Mu2e experiment is to improve the $\mu^-N \rightarrow e^-N$ search by four orders of magnitude.
  
There is no observable Standard Model contribution to the $\mu^-N \rightarrow e^-N$ signal at Mu2e.   Neutrino oscillations imply that muon-to-electron conversion can proceed via a penguin diagram that contains a $W$ and an oscillating neutrino. However, the rate for this conversion process is more than 30 orders of magnitude
below the projected sensitivity of the Mu2e experiment.  Any signal observed at Mu2e would be an unambiguous indication of new physics.

Many Standard Model (SM) extensions~\cite{Marciano:2008zz} include charged-lepton-flavor-violating interactions that would force this process to occur at a rate much higher than the one expected in the SM.  In fact, many new physics models predict a conversion rate to which Mu2e would have discovery sensitivity.  Examples include supersymmetry with and without $R$-parity conservation, models with multiple Higgs doublets, $Z'$ models, leptoquark models, model of extra dimensions, etc.~\cite{review,Kuno:1999jp,Calibbi:2006nq}.

\begin{figure}
\centering
\includegraphics[width=0.6\textwidth]{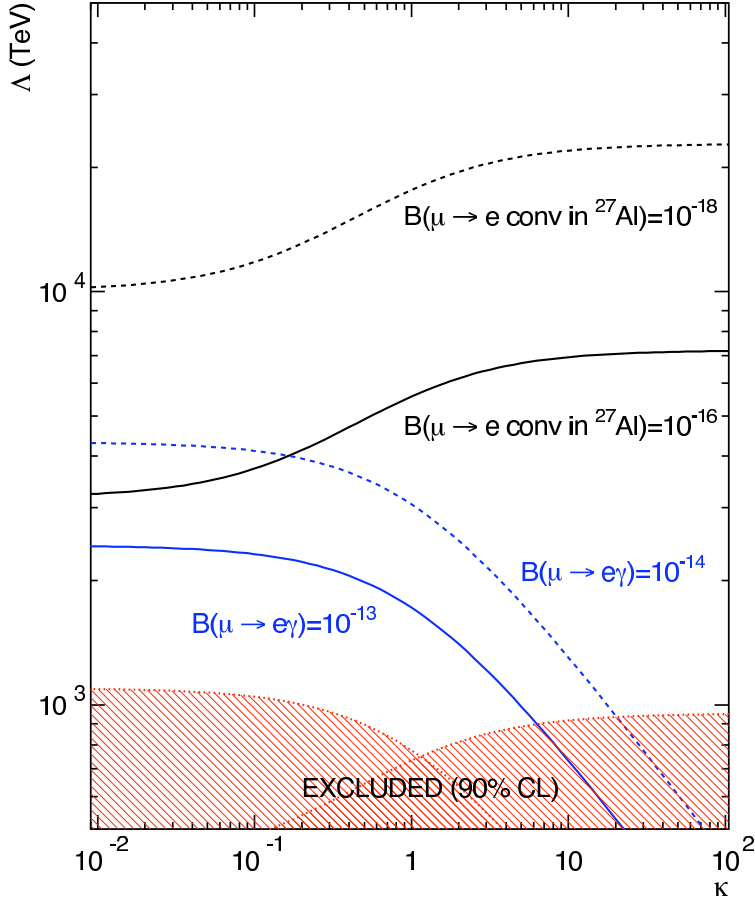}
\caption{Physics reach of muon lepton-flavor-violation searches.
The shaded region on the left is excluded by MEG~\cite{Adam:2013mnn}, and on the right by SINDRUM~II~\cite{Bertl:2006up}.  The figure is from~\cite{deGouvea:2013zba}.}
\label{fig:Lambda-kappa}
\end{figure}

Depending on the model, CLFV interactions can be mediated via a loop or a four-fermion contact interaction.    Without specifying the model for new physics, an effective-theory Lagrangian containing both terms can be written as follows:
 \begin{equation}
 \frac{\ds{}m_{\mu}}{\ds(1+\lfvkappa)\,\lfvlambda^2}\,{\bar{\mu}}_{R}\sigma_{\mu\nu}e_{L}F^{\mu\nu}
+
\frac{\ds\lfvkappa}{\ds{}(1+\lfvkappa)\,\lfvlambda^2}\,{\bar{\mu}}_L\gamma_{\mu}e_L
  ( {\bar{u}}_L\gamma^{\mu}u_L +  {\bar{d}}_L\gamma^{\mu}d_L  )
\end{equation}
where $\lfvlambda$ is the mass scale of the new physics, and the
dimensionless parameter $\lfvkappa$ represents the
relative strength of the contact term.  $\lfvkappa$ is used to illustrate the physics
reach of muon lepton-flavor-violating searches in
Fig.~\ref{fig:Lambda-kappa} taken from Ref.~\cite{deGouvea:2013zba}.  The first term in the equation above typically generates magnetic-moment loop diagrams where a photon is exchanged.

The planned discovery sensitivity goal of the Mu2e experiment is approximately the
$10^{-16}$ conversion rate line on Fig.~\ref{fig:Lambda-kappa}.
From the figure, it is clear that the Mu2e experiment will provide an improvement over a broad range of models, independent of $\lfvkappa$,
and will be sensitive to mass scales ($\lfvlambda$) of thousands of TeV.    
Also clear from the figure is that searches for $\mu^-N \rightarrow e^-N$ and $\mu \rightarrow e\gamma$ are complementary.  If a signal of charged-lepton-flavor violation is observed, then the relative rates of $\mu \rightarrow e\gamma$,  $\mu^+ \rightarrow e^+e^-e^+$, and the coherent $\mu^-N \rightarrow e^-N$ conversion processes can be used to constrain the underlying physics responsible for the observed CLFV interactions.  

Searches for charged-lepton-flavor violation are also complementary to new physics searches at the Large Hadron Collider.  In many scenarios of new physics observation at the Large Hadron Collider it is difficult to determine if lepton flavor is violated.  As an example, many scenarios that predict a SUSY signal within reach of the Large Hadron Collider also predict a value for $R_{\mu e}\approx 10^{-15}$ for which Mu2e would observe about 50 events on a background of less than one event.  In addition, if the experiments at the Large Hadron Collider do not find a signal of new physics, Mu2e will probe energy scales up to thousands of TeV, so, it may be our best chance to probe physics beyond the standard model.

\subsection{The Mu2e Experiment at Fermilab}
\label{sec:Mu2e}

The Mu2e experiment will search for the conversion of a muon into an electron in the presence of a
nucleus~\cite{Mu2eCDR} at a sensitivity of about four orders of magnitude beyond the best current limits.  The observation of this process would be a major discovery, signaling the existence of charged-lepton-flavor violation far beyond what is expected from current standard
theory.  A non-observation would be equally interesting as it would place stringent limits on theory and exclude large regions of parameter space for leading theories of beyond-standard-model physics. 
The Mu2e experiment obtained DOE CD-1 approval in June 2012, is scheduled to obtain DOE CD-2 approval in mid-2014, and is scheduled to begin operations in 2019. 

\begin{figure}
\centering
\includegraphics[width=0.8\textwidth]{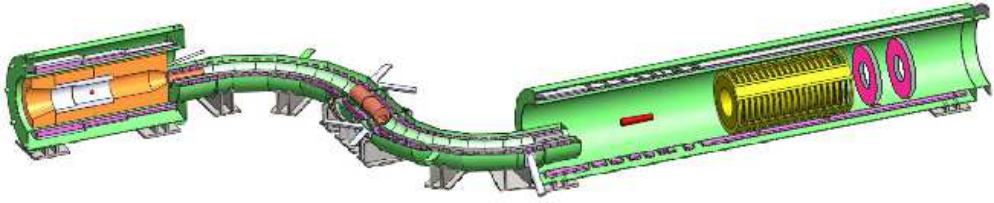}
\caption{The Mu2e experimental setup. The pulsed proton beam enters the production solenoid (far left) from the top right.  The muons produced are captured by the production solenoid and transported through the S-shaped transport solenoid to the aluminum stopping target (small red cylinder).  Electrons produced in the stopping target are captured by the magnetic field of the detector solenoid (right) and transported through the tracker (yellow) where the momentum is measured.  The electrons then strike the electromagnetic calorimeter (pink annuli), which provides an independent measurement of the electron energy.  A cosmic-ray-veto system and some other parts of the apparatus are not shown.}
\label{fig:mu2e}
\end{figure}

The approach of the Mu2e experiment is to stop low-momentum muons from a pulsed beam on an aluminum target to form muonic atoms and then to measure the resulting electron spectrum.  The signal would produce a mono-energetic electron with an energy of about 105~MeV.  In order to reach the design sensitivity (single-event sensitivity of $2\times 10^{-17}$), about $10^{18}$ muons must be stopped.  Keeping the background expectation to less than one event in this high-intensity experiment is obviously quite a challenge and results in the unique experimental setup summarized below and depicted in Fig.~\ref{fig:mu2e}.

The first step in the experiment is to produce the low-momentum pulsed muon beam. Existing Tevatron infrastructure will be repurposed to deliver 8~GeV protons with 1695~ns bunch spacing to a tungsten production target. About $3\times10^7$ protons per bunch will be delivered with a duty factor of about 30\%.  The 1695~ns bunch spacing is well suited for the Mu2e experiment given that the lifetime of the muonic aluminum is about 864~ns.   The resulting pions and muons produced inside the production solenoid are collected and passed to the S-shaped muon beamline where absorbers and collimators are optimized to eliminate positive particles and anti-protons while efficiently transmitting low-energy negatively-charge pions and muons.  Most of the pions will decay inside the 13~m long beamline, while about 40\% of surviving muons will be stopped in an aluminum stopping target.  Simulations estimate that Mu2e will produce $0.0016$ stopped muons per proton on target.

Muons stopped in the target will drop to the K-shell, forming a muonic atom.  As muons settle into the K-shell a cascade of X-rays will be emitted.  By measuring these X-rays the rate of stopped muons can be established from which the denominator of $R_{\mu e}$ can be determined. About 60\% of stopped muons will undergo muon capture on the nucleus while the other 40\% will decay in orbit (DIO).  The DIO process produces an electron with a continuous Michel distribution including a long tail due to photon exchange with the nucleus.  In the limit where the neutrinos carry no energy from the Michel decay, the electron carries the maximum energy of 105~MeV.  In this limit, the DIO electron is indistinguishable from the $\mu^-N \rightarrow e^-N$ conversion signal.  In addition, mis-measurements of the DIO electron momentum contribute to an irreducible background.  In order to combat the DIO background, the Mu2e experiment requires a tracking detector with momentum resolution of about $0.1$\%.    

To accomplish the required momentum resolution, the Mu2e tracker will use straw tubes in vacuum.  The inner radius of the tracker is empty so that only tracks with
transverse momenta above $53$~MeV/c will pass through any of the straw tubes. The Mu2e scintillating-crystal calorimeter will provide cross checks of signal candidates and particle identification.  The calorimeter also has an empty inner-radius region.  The empty inner regions effectively make the tracker and calorimeter blind to the bulk of the DIO electrons and to muons that don't stop in the aluminum target and thus allow these detectors to handle the high-intensity environment required for the experiment.    

The primary backgrounds for the Mu2e experiment can be classified into several categories: intrinsic muon-induced backgrounds, prompt backgrounds, and miscellaneous backgrounds, primarily anti-proton induced and cosmic-ray induced backgrounds. The intrinsic backgrounds arise from muon DIO and radiative muon capture.  The dominant prompt background arises from radiative pion capture and subsequent conversion of the $\gamma$ in the stopping target material to produce a 105~MeV electron.  Prompt backgrounds like this can be controlled by taking advantage of the long muon lifetime and by optimizing the properties of the pulsed beam.  After a proton pulse there is a delay of about 700~ns before Mu2e begins accepting candidate signal events.  This delayed signal time window reduces the pion-related backgrounds by about 10 orders of magnitude.  In order to avoid prompt backgrounds that arise from out-of-time protons that arrive at the production target near the start of the signal time window Mu2e requires the ratio of out-of-time protons to in-time protons be less than $10^{-10}$.  This suppression of out-of-time protons will be achieved and monitored with dedicated systems. Beam electrons, muon decay in flight, and pion decay in flight are other prompt backgrounds that are defeated through the combination of the pulsed beam and delayed signal window.  

As mentioned above muon DIO is an intrinsic background with an electron spectrum that extends up to the signal region at the kinematic end
point~\cite{czarnecki}.  Another intrinsic background is due to radiative muon capture on the target foils producing a high-energy photon that can convert into an electron-positron pair.  The kinematic endpoint of the electron energy distribution is slightly below 105 MeV, so this background, like the DIO, can be mitigated by minimizing non-Gaussian contributions to the tails of the momentum resolution.

Cosmic-ray muons may interact within the detector solenoid region and produce background electrons.  Passive shielding and an active cosmic-ray-veto system are employed to ensure that cosmic rays are a subdominant background.

The background expectation in the Mu2e experiment for a  three-year run at 8~kW beam power is summarized in Table~\ref{tab:Mu2eBackgrounds}.  

\begin{table}[t]
  \centering
  \begin{tabular}{llcr} \hline\hline
    Category      
         & Source    &\hspace{0.15in}   & Events \\ \hline
    \multirow{2}{*}{Intrinsic}    
         & $\mu$ decay in orbit       & & 0.22   \\ 
         & radiative $\mu$ capture    & & $<0.01$  \\ \hline
    \multirow{4}{*}{Prompt}
         & radiative $\pi$ capture    & & 0.03   \\
         & beam electrons             & & $<0.01$  \\
         & $\mu$ decay in flight      & & 0.01   \\
         & $\pi$ decay in flight      & & $<0.01$  \\ \hline
    \multirow{3}{*}{Miscellaneous}              
         & anti-proton induced        & & 0.10   \\
         & cosmic-ray induced         & & 0.05   \\
         & pat. recognition errors    & & $<0.01$  \\ \hline
    Total Background 
         &                            & & 0.41   \\ \hline\hline
  \end{tabular}
  \caption{A summary of the current Mu2e background estimates.
  }
  \label{tab:Mu2eBackgrounds}
\end{table}

\subsection{A Second-Generation Mu2e Experiment at Fermilab}
\label{sec:Mu2e2}

The current Mu2e experiment is being designed to have a single-event sensitivity of $\sn{2}{-17}$ with a total background of $< 0.5$ events to reach a target sensitivity of $R_{\mu e} < 6\times 10^{-17}$~\cite{Mu2eCDR} at the 90\% confidence level.  To reach that sensitivity, about $10^{18}$ stopped muons are collected from pions produced by an 8 GeV proton beam colliding with a tungsten target.  The proton beam is pulsed with a 1695~\ns\ spacing (center-to-center) and the proton pulses have a full width of $\pm 100$~\ns.  An aluminum stopping target is used. To suppress prompt backgrounds from beam electrons and from radiative pion captures (RPC) the signal timing window is delayed about $700$~\ns\ relative to the center of the proton pulse.  To suppress radiative pion capture background from pions that originate from protons arriving late at the tungsten production target the extinction -- defined as the ratio of the number of out-of-time protons to the number of in-time protons -- must be $10^{-10}$.  Backgrounds that result from anti-proton annihilations near the aluminum stopping target are mitigated by placing a thin beryllium window half-way along the transport solenoid.  This window annihilates the anti-protons far from the stopping target while reducing the stopped-muon yield by about 10\%. Some relevant beam parameters are given in Table~\ref{tab:Mu2eBeamParameters} while the background contributions are summarized in Table~\ref{tab:Mu2eBackgrounds}.  These are reproduced from reference~\cite{Mu2eCDR}.

Here, we explore the opportunity and limitations of upgrading the current Mu2e experiment described above.  The relevant beam parameters are given in Table~\ref{tab:Mu2eBeamParameters} for Mu2e and for the Project-X beam upgrades considered here.  Our goal is to determine if it is feasible that a Mu2e-II, based largely on the Mu2e apparatus and using Project-X beams, can achieve a single-event sensitivity of $\sn{2}{-18}$ while keeping the total expected background small.  We use simulation to estimate the muon and pion stopping rates, the muon and pion arrival time (at the stopping target) distributions, and the muon decay time distribution.  With these rates and distributions in hand we then estimate how the backgrounds will scale for various Mu2e-II scenarios.  We explore aluminum, titanium, and gold stopping targets for Project-X beams at 1~\gev\ and 3~\gev.  Finally, we discuss the limitations of the Mu2e facility at higher beam intensity.    

\begin{table}[tb]
  \centering
  \begin{tabular}{rccc} 
  \hline\hline
    &\hspace*{0.15in} & Mu2e  & Project-X \\ \hline
  POT                     & & \sn{3.6}{20}     & varies by scenario \\
  extinction              & & $< \sn{1}{-10}$  & $< \sn{1}{-12}$ \\
  run duration (years)    & & 3                & 3 \\
  run time (sec/year)     & & \sn{2}{7}        & \sn{2}{7} \\
  duty factor             & & 0.30             & 0.90 \\
  p pulse full width (ns) & & 200              & 100 \\
  p pulse spacing (ns)    & & 1695             & 1695 \\
  signal start time (ns)  & & 670              & 670 \\
  signal stop time (ns)   & & 1595             & 1645 \\
  live gate fraction      & & 0.55             & 0.58 \\
  beam energy (GeV)       & & 8                & 1 or 3 \\     
  beam power (kW)         & & 8                & varies by scenario \\ 
  \hline\hline
  \end{tabular}
  \caption{Some relevant beam parameters for the currently planned Mu2e 
    and those assumed for some potential Project-X scenarios.
  }
  \label{tab:Mu2eBeamParameters}
\end{table}

\section{Simulation}
\label{sec:simulation}

To estimate the rates and distributions discussed in the previous section, we use \texttt{G4Beamline v2\_12} as
developed at Muons, Inc.~\cite{g4bl}, which is a simplified version
of Geant4, primarily intended for beamline studies.  Three sets of
simulated experiments are run based on the parameters shown in
Table~\ref{tab:energies}.  The 8~\gev case corresponds to the current
Mu2e configuration.  For potential Project-X upgrades, we consider 1 and 3~\gev\ kinetic energy protons in pulses with a full width of $\pm50$~\ns.  In all instances the full Mu2e solenoid system is described including all collimators, the production solenoid heat-and-radiation shield, the anti-proton window, and the latest magnetic-field map.  The stopping-target geometry is described in~\cite{Mu2eCDR} and is left unchanged for the different scenarios.

\begin{table}[b]
  \centering
  \begin{tabular}{lcccc} \hline\hline
     & Kinetic Energy  & Momentum & Simulated Events\\ \hline
    Baseline Mu2e  & 8 \gev & 8.889 \gevc & \sn{100}{6} \\
    Project-X Mu2e & 3 \gev & 3.825 \gevc & \sn{200}{6} \\
    Project-X Mu2e & 1 \gev & 1.696 \gevc & \sn{500}{6} \\ \hline\hline
  \end{tabular}
  \caption{Proton beam parameters considered in this study.  The assumed proton mass is 938.272 \mev.  The 8~\gev\ row corresponds to the current Mu2e parameters while the 1 and 3~\gev\ rows correspond to expected parameters at Stage-1 and Stage-2 of Project-X, respectively.}
  \label{tab:energies}
\end{table}

Since charged pions have a proper lifetime of $\tau = 26~\ns$,
the probability for them to survive to the stopping target is very
small.  Running enough simulations to get an accurate timing
distribution to the precision we need is therefore impractical.  To
resolve this complication, we set the charged pion lifetime to
effectively infinity, and then weight the resulting stopping-time
distribution according to the survival probability:
%
$P(t,\gamma) = \exp\left(-\frac{t}{\gamma\tau}\right)$
%
where $t$ is the stopping time of the pion in the laboratory frame,
and $\gamma$ is the time dilation factor given by
%
$\gamma = \sqrt{\left(\frac{pc}{m_\pi c^2}\right)^2 + 1}$,
%
where $p$ is the momentum of the particle as it enters the detector
solenoid (DS).  This prescription is not precisely correct as $\gamma$
is not likely to remain constant throughout the pion's transit from
the production target to the stopping target.  We expect such an
effect to influence results at no more than the few-percent level,
and thus unlikely to significantly affect the conclusions of the current study.

%

\subsection{Proton pulse shape}
\label{sec:pot}

The timing distribution of the proton pulse in \texttt{G4Beamline} is modeled as a delta function located at $t = 0~\ns$.  In order to get a more accurate estimate of the experimental sensitivity we convolute the relevant timing distributions with the expected shape of the proton pulse as estimated using dedicated simulations of the Mu2e proton beam.  

\begin{figure}[bt]
  \centering
  \includegraphics[width=0.75\textwidth]{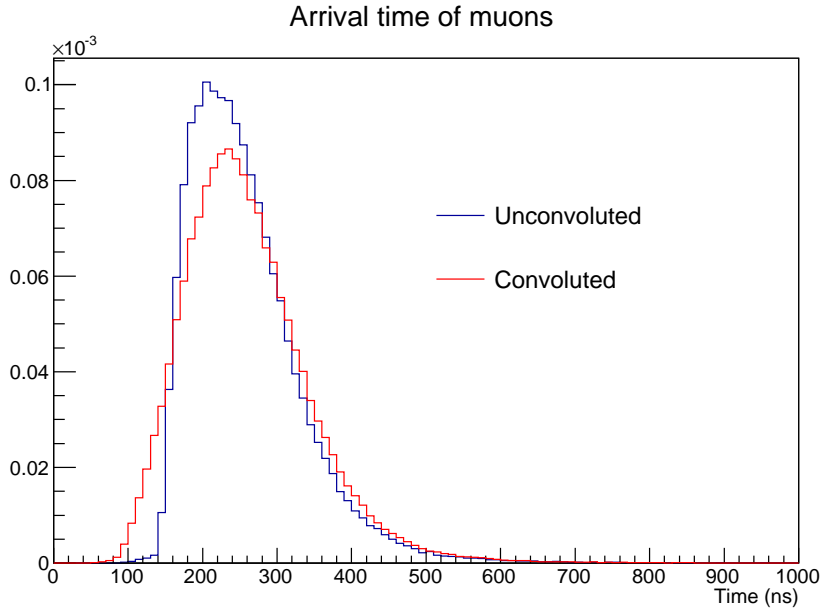}
  \caption{Arrival time of stopped muons on the aluminum stopping target 
    before (blue) and after convolution (red) with the proton pulse
    shape estimated for Mu2e.  Along the horizontal axis, $t=0$ 
    corresponds to the center of the proton pulse as it arrives at the 
    tungsten production target.}
  \label{fig:conv_muons}
\end{figure}

\begin{figure}[bt]
  \centering
  \includegraphics[width=0.49\textwidth]{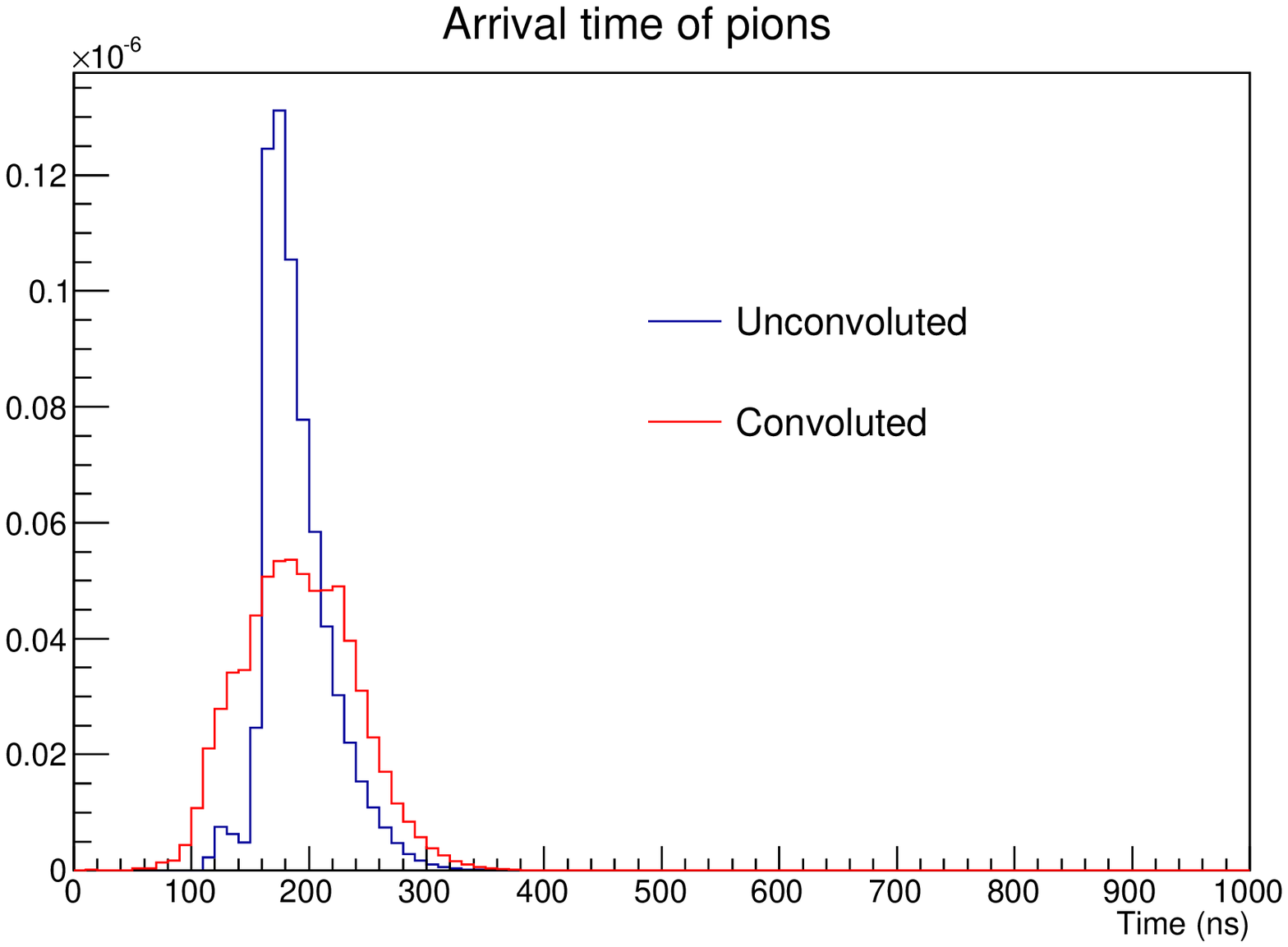} \hfill
  \includegraphics[width=0.49\textwidth]{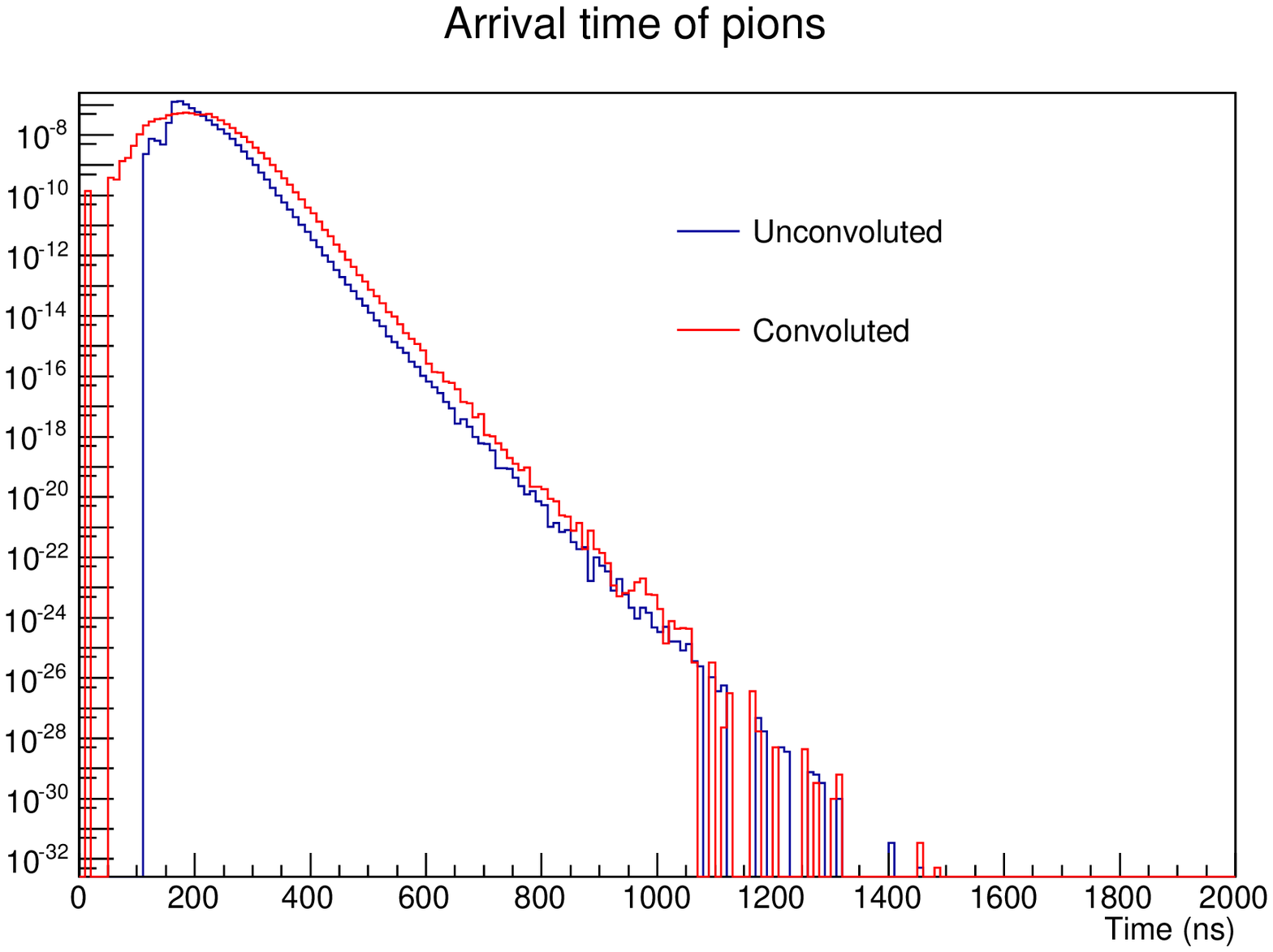}
  \caption{Arrival time of stopped pions, weighted by their survival
    probability, on the aluminum stopping target 
    before (blue) and after convolution (red) with the proton pulse
    shape simulated for the Mu2e experiment. The left plot has a linear 
    scale while the right plot has a logarithmic scale and an extended 
    horizontal axis.}
  \label{fig:conv_pions}
\end{figure}

The effects of including the proton pulse shape on the arrival time of stopped muons and stopped pions are
shown in Fig.~\ref{fig:conv_muons} and Fig.~\ref{fig:conv_pions} for
the Mu2e project.  For Project-X
the proton pulses are expected to be significantly narrower, $\pm 50$~ns for 100 kW of beam power~\cite{bobT}.  For the Project-X studies we assume the same proton pulse shape as supplied for Mu2e, but reduce the width of the pulse by a factor of two.

\begin{figure}[bt]
  \centering
  \includegraphics[width=0.49\textwidth]{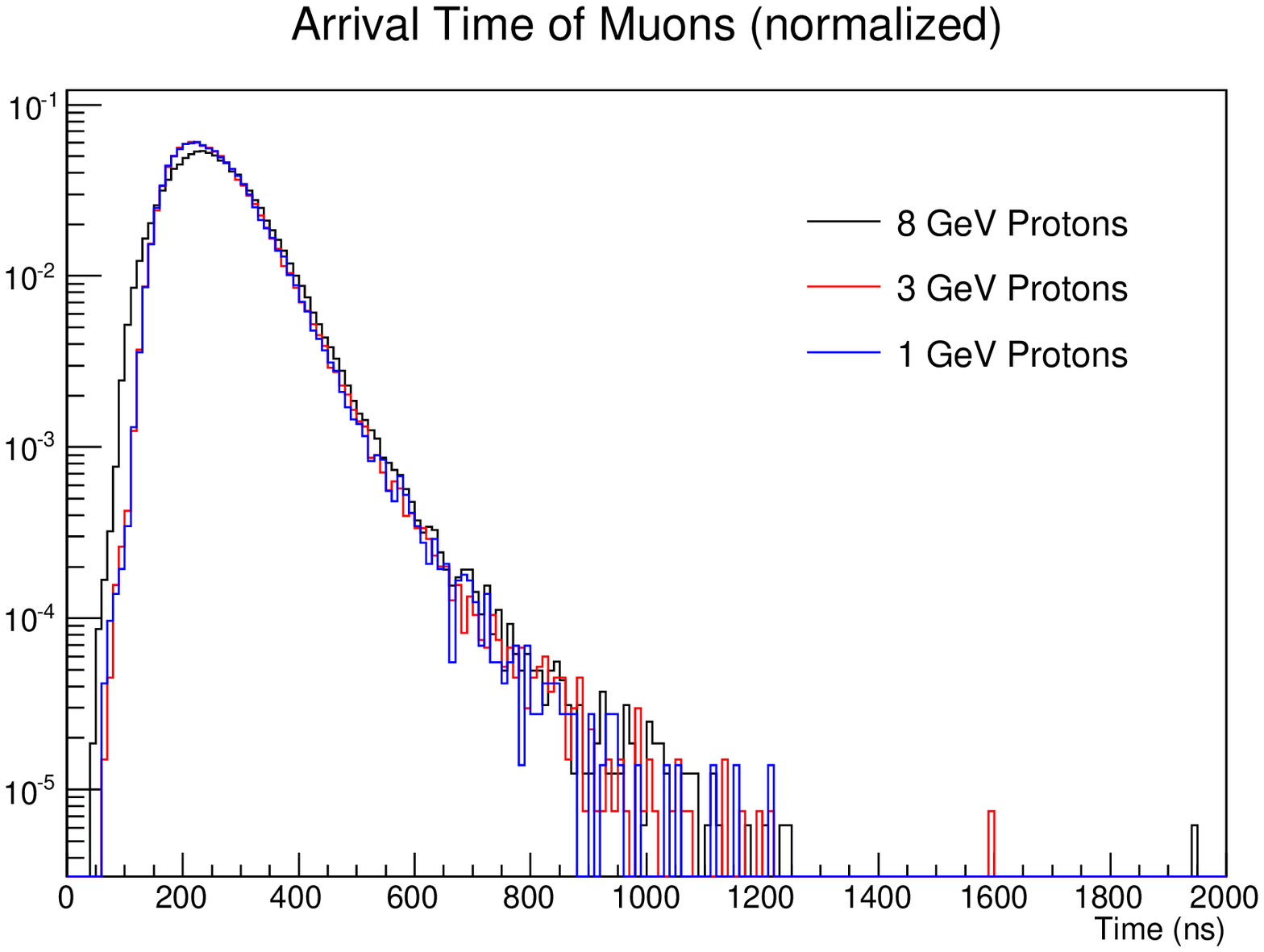}\hfill
  \includegraphics[width=0.49\textwidth]{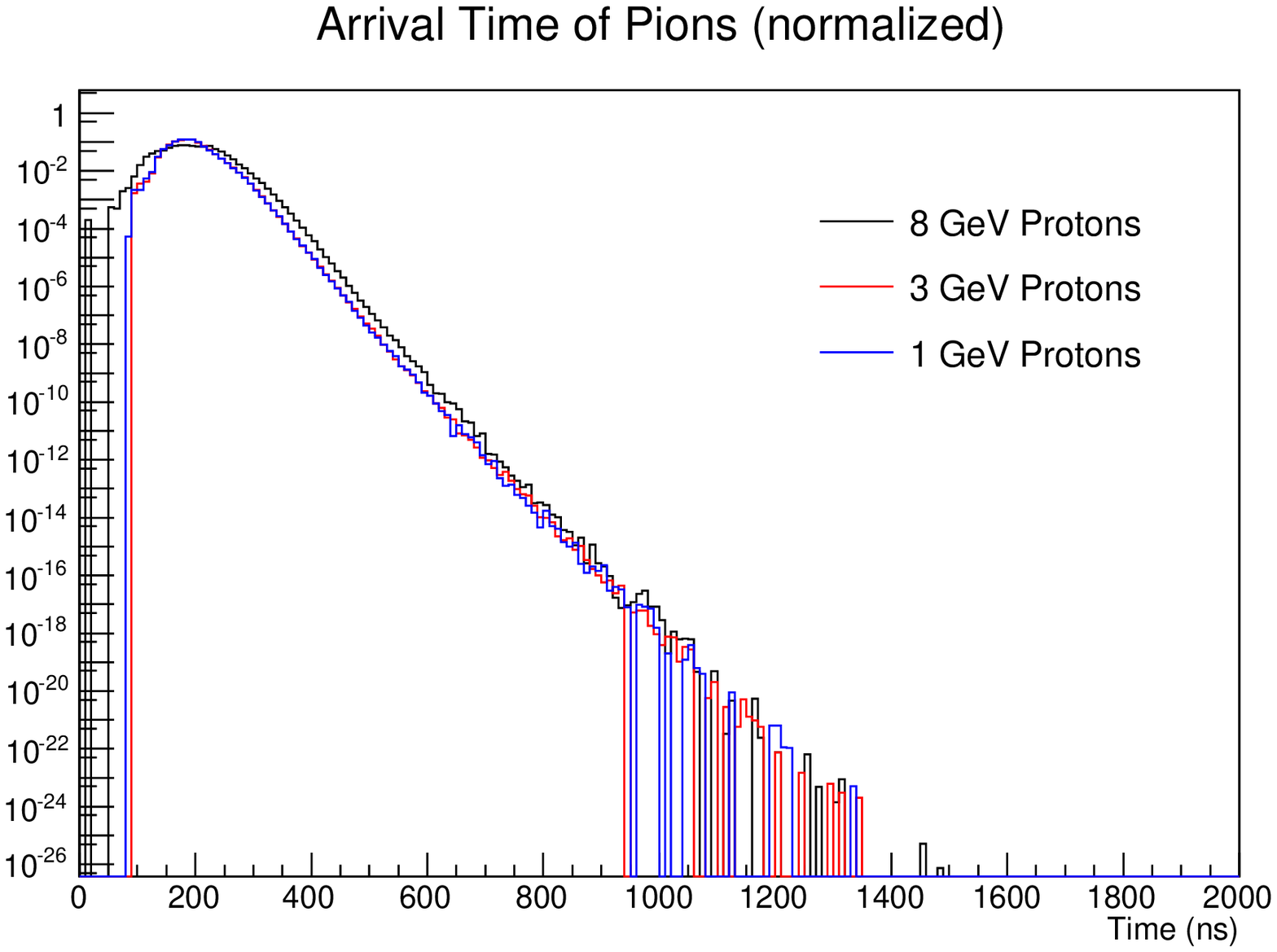}
  \caption{Arrival time of stopped muons (left) and stopped pions (right)
    for the Project-X scenario for 1 and 3~\gev\ protons after convolution
    with the proton pulse shape.  The pion distribution has been weighted
    by the pion survival probability.  The histograms have been normalized
    to the same area in each plot separately.
  }
  \label{fig:PXstoptimes}
\end{figure}

We follow this same procedure to obtain the muon and pion stopping-time distributions for Project-X beam at 1 and 3~\gev.  These plots are shown in Fig.~\ref{fig:PXstoptimes} after convolution with the proton pulse shape and weighting by the pion survival probability.  The stopping-time distributions are largely independent of the stopping target material and for a given beam configuration we use the same stopping-time distributions for each of the stopping target nuclei investigated.

\FloatBarrier
\subsection{Simulation results}
\label{sec:SimResults}

\begin{figure}[tb]
  \centering
  \includegraphics[width=\textwidth]{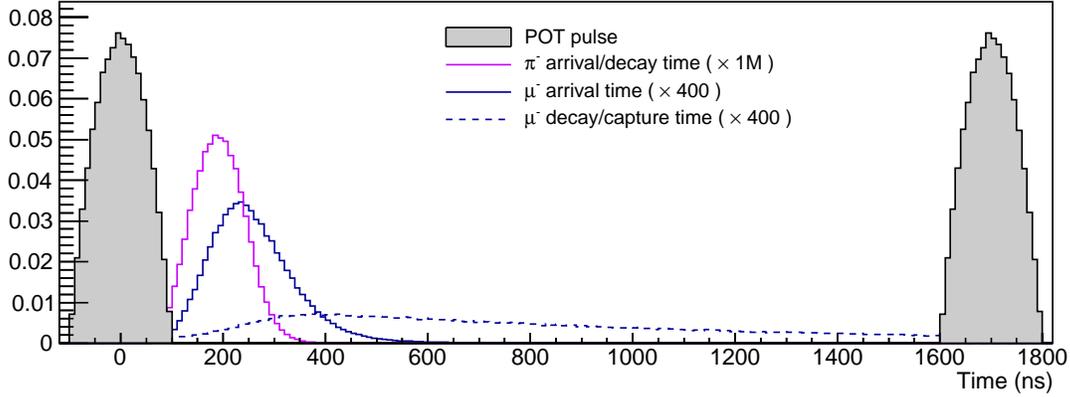}
  \caption{Timing distribution for the Mu2e experiment.  This
  shows a figurative proton pulse width, the arrival time of the charged
  pions and muons, and the capture-time distribution of the muons on
  an aluminum stopping target.}
  \label{fig:timebaseline}
\end{figure}

The overall timing distribution for the Mu2e experiment is shown in Fig.~\ref{fig:timebaseline} for an aluminum stopping target. Since the $\pi^-$-N interaction is strong, the pion is captured by the nucleus nearly instantaneously relative to the pion stopping time.  The pion capture-time distribution is therefore assumed to be the same as the pion arrival-time distribution.  For muons, the capture rate is characterized by a falling exponential, $\exp^{-t/\lambda}$, where $\lambda$ is the effective lifetime of the bound muon and is nuclei dependent.  In addition, the fractions of bound muons that are captured or that decay-in-orbit are also nuclei dependent.  These nuclei-dependent characteristics affect the sensitivity of a given experiment and are listed in Table~\ref{tab:nucleistuff} for the stopping-target nuclei we considered.

\begin{table}[tb]
  \centering
  \begin{tabular}{rccc} \hline\hline
           & lifetime (ns) & capture fraction & decay fraction \\ \hline
  Aluminum & 864           & 0.61            & 0.39 \\
  Titanium & 297           & 0.85            & 0.15 \\ \hline\hline
  \end{tabular}
  \caption{The lifetime of the bound muon and the muon capture and decay     
    fractions for various stopping target nuclei that affect the 
    sensitivity estimates for Mu2e-II. 
  }
  \label{tab:nucleistuff}
\end{table}

The muon capture-time distribution is shown in Fig.~\ref{fig:timebaseline} as the blue dashed line.  In this figure and all subsequent figures the timing distributions are folded over modulo 1695~\ns\ in order to account for contributions from previous protons pulses. Quantities associated with these curves are shown in Table~\ref{tab:8gev}.  Note that for presentation purposes the proton pulse distribution shown in the figure is a simplified version of the actual distribution.  All the rates and numbers in the tables are derived using the actual distribution without simplification.

\begin{table}[tb] 
  \centering
  \begin{tabular}{lcc}\hline\hline
                                          & Muons         & Pions         \\  \hline
    Stops/POT                             & \sn{1.61}{-3} & \sn{6.82}{-7} \\
    Stop Time (Mean)                      & 259.8 ns      & 192.1 ns      \\
    Stop Time (rms)                       & 86.58 ns      & 47.6 ns       \\
    Fraction of $\mu$-captures in live gate & 0.489         & ---           \\
    Fraction of $\pi$-captures in live gate &  ---          & \sn{3.93}{-11}\\
    \hline\hline
  \end{tabular}
  \caption{Stopped muon and pion quantities for the Mu2e
  experiment.  The live gate is defined by $670 < t < 1595$ ns.
  The fraction of captures includes contributions from the
  previous proton bunch.}
  \label{tab:8gev}
\end{table}

\begin{figure}[tb]
  \centering
  \includegraphics[width=\textwidth]{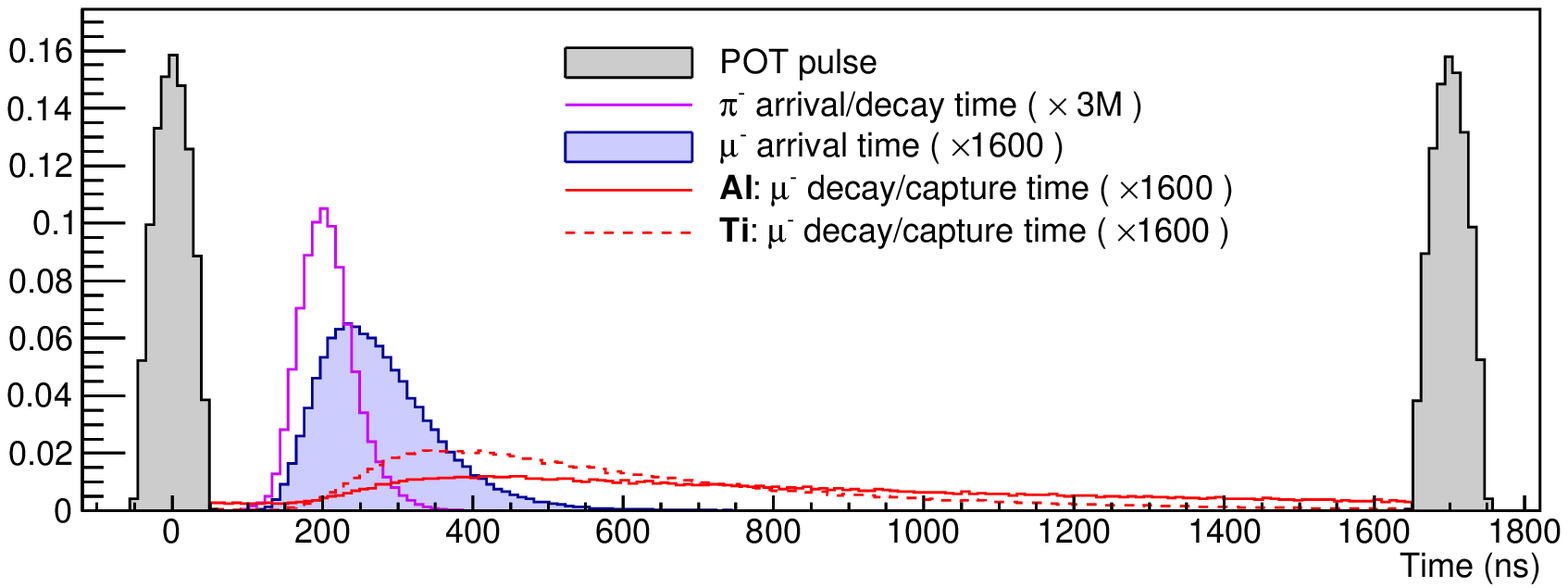}
  \caption{Timing distribution for the 3~\gev\ Project-X case.  Shown
  here is a figurative POT pulse width, the arrival time of the
  charged pions and muons, and the capture-time distribution of the
  muons on aluminum and titanium stopping targets.}
  \label{fig:3gev}
\end{figure}

Figure~\ref{fig:3gev} shows the Project-X 3~\gev\ scenario, where the proton pulse width is half that of the 8~\gev\ configuration, and both aluminum and titanium are considered as a stopping target. Because the proton pulse width is narrower, we can lengthen the live gate~\cite{MoveLiveGate}. The associated quantities are shown in Table~\ref{tab:3gev}.  Note that the arrival times for stopped particles are slightly earlier for titanium than for aluminum (not depicted in the figure, which just shows the difference in capture times on the stopping target). This is because titanium is heavier and can therefore stop the more energetic muons/pions that are not stopped by the aluminum target.

The same exercise was performed for the Project-X 1~\gev\ scenario, the
results of which are presented in Table~\ref{tab:1gev}.  With the
exception of the stopping yields, the parameters of the timing
distributions for the 1~\gev\ case are very similar to those of the
3~\gev\ case.

\begin{table}[tb] 
  \centering
  \begin{tabular}{lccccc}\hline\hline
                     & \multicolumn{2}{c}{Aluminum target} &\hspace*{0.15in}       & \multicolumn{2}{c}{Titanium target} \\
                     & Muons              & Pions                 & & Muons         & Pions               \\  \hline
    Stops/POT        & \sn{6.69}{-4}      & \sn{2.90}{-7}         & & \sn{8.68}{-4} & \sn{6.35}{-7}       \\
    Stop Time (Mean) & 257.8 ns           & 190.1 ns              & & 242.6 ns      & 177.1 ns            \\
    Stop Time (rms)  & 79.03 ns           & 34.5 ns               & & 77.34 ns      & 33.7 ns             \\
    Fraction of $\mu$-captures in live gate & 0.497 & ---           & & 0.269         & ---                 \\
    Fraction of $\pi$-captures in live gate &  ---  & \sn{1.14}{-11} & &  ---          & \sn{4.74}{-12}      \\
    \hline\hline
  \end{tabular}
  \caption{Stopped muon and pion quantities for a Project-X scenario
  with a narrower pulse width and 3~\gev\ protons.  The live gate here
  is defined by $670 < t < 1645$~\ns.  The fraction of captures
  includes contributions from the previous proton bunch.}
  \label{tab:3gev}
\end{table}

\begin{table}[tb] 
  \centering
  \begin{tabular}{lccccc}\hline\hline
                     & \multicolumn{2}{c}{Aluminum target} &\hspace*{0.15in} & \multicolumn{2}{c}{Titanium target} \\
                     & Muons         & Pions                      & & Muons         & Pions               \\  \hline
    Stops/POT        & \sn{1.44}{-4} & \sn{6.44}{-8}              & & \sn{1.87}{-4} & \sn{1.40}{-7}       \\
    Stop Time (Mean) & 257.4 ns      & 189.7 ns                   & & 242.3 ns      & 176.7 ns            \\
    Stop Time (rms)  & 78.46 ns      & 34.3 ns                    & & 76.83 ns      & 33.5 ns             \\
    Fraction of $\mu$-captures in live gate & 0.496 & ---           & & 0.266         & ---                 \\
    Fraction of $\pi$-captures in live gate &  ---  & \sn{1.40}{-11} & &  ---          & \sn{4.51}{-12}      \\
    \hline\hline
  \end{tabular}
  \caption{Stopped muon and pion quantities for a Project-X scenario
  with a narrower pulse width and 1~\gev\ protons.  The live gate here
  is defined by $670 < t < 1645$ ns.  The fraction of captures
  includes contributions from the previous proton bunch.}
  \label{tab:1gev}
\end{table}

%

\begin{figure}[htb]
  \centering
  \includegraphics[width=\textwidth]{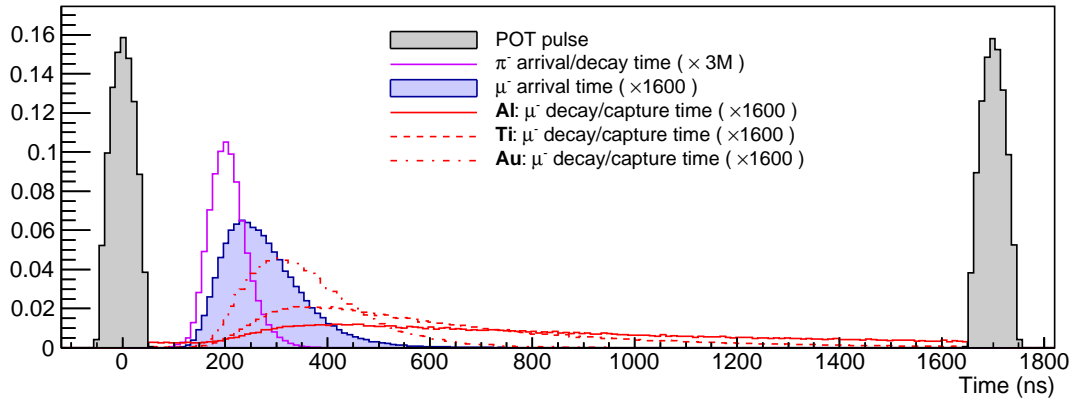}
  \caption{Timing distribution for the 3-GeV Project-X case.  Shown
  here is a figurative POT pulse width, the arrival time of the
  charged pions and muons, and the capture-time distribution of the
  muons on aluminum, titanium, and gold.}
  \label{fig:3gev_Au}
\end{figure}

We also produced the muon capture-time distribution for a gold stopping target as shown in Fig.~\ref{fig:3gev_Au}.  Since the lifetime of the bound muon in gold is so small (72~ns) the fraction of muon captures that occur within the live gate is quite small -- only 1.22\% for a live gate of $670 < t < 1645~\ns$.  Since the muon capture-time distribution has such a large overlap with the pion arrival time distribution, it's clear that it will be difficult to achieve a reasonable signal acceptance while sufficiently suppressing the radiative-pion-capture background.  To achieve the necessary pion/muon separation requires a dedicated study of alternative transport systems and is beyond the scope of this study. We do not further consider the gold stopping target.

\FloatBarrier
\section{Estimating backgrounds for Mu2e-II scenarios}
\label{sec:bgd}

We use the quantities in Tables~\ref{tab:8gev}, \ref{tab:3gev}, and \ref{tab:1gev} to estimate the backgrounds for a variety of Mu2e-II scenarios.  For most of the background sources we estimate the Mu2e-II background yields by scaling from the Mu2e yields in Table~\ref{tab:Mu2eBackgrounds}, which were derived using full simulation as described in~\cite{Mu2eCDR} and references therein.  Since it is especially sensitive to tails in the momentum resolution, full simulations - including the effects of the additional occupancy expected for Mu2e-II - are performed to estimate the DIO background yield.

In Tables~\ref{tab:3gev} and~\ref{tab:1gev} the stopping rates are larger for titanium than for aluminum due to the fact that in the simulations we left the stopping-target geometry unchanged and that titanium is heavier and at a larger $Z$.  In that configuration a titanium stopping target would also have a poorer momentum resolution since scattering in the stopping target is a significant contribution to the spectrometer resolution.  Dedicated simulations show that if the titanium stopping-target mass is reduced so that it gives the same muon-stopping yields as aluminum, it then has the same momentum resolution as the aluminum stopping target. Thus, in the calculations below we always use the aluminum stopping yields for a given proton beam energy and scale from the Mu2e background estimates assuming that the Mu2e-II reconstruction efficiencies and resolutions are unchanged relative to those of Mu2e.  We comment on the validity of this assumption in Sec.~\ref{sec:rates} below.

\subsection{Necessary number of protons on target}
\label{sec:npot}

We begin by estimating the total number of protons on target required in order to improve the sensitivity by a factor of ten.  We start from the current Mu2e estimates and apply the following scaling:
\begin{equation}
\label{eq:npot}
 N_{\mathrm{POT}}^{(A,Z)} = 10\cdot N_{\mathrm{POT}}^{\mathrm{Mu2e}}
  \cdot \left( \frac{S^{\mathrm{Mu2e}}} {S^{(A,Z)}_{E(p )}} \right)
  \cdot \left( \frac{C^{\mathrm{Mu2e}}} {C^{(A,Z)}} \right)
  \cdot \left( \frac{F^{\mathrm{Mu2e}}} {F^{(A,Z)}} \right)
  \cdot \left( \frac{\varepsilon^{\mathrm{Mu2e}}} {\varepsilon^{(A,Z)}} \right),
\end{equation} 
where $N_{\mathrm{POT}}^{(A,Z)}$ is the total number of POT necessary for Mu2e-II to achieve a sensitivity that is $\times 10$ smaller than Mu2e for a given stopping-target nuclei $(A,Z)$, $S^{(A,Z)}_{E(p )}$ is the number of stopped muons per POT for a given target nuclei with protons of energy $E(p )$, $C^{(A,Z)}$ is the muon capture fraction for a given stopping-target nuclei, $F^{(A,Z)}$ is the fraction of muon captures that occur in the signal-time window for a given stopping-target nuclei, and $\varepsilon^{(A,Z)}$ is the total reconstruction and selection efficiency for electrons from $\mu^- N\rightarrow e^- N$ interactions. The superscript $\mathrm{Mu2e}$ denotes those quantities corresponding to the current Mu2e design, which uses an aluminum stopping target ($\mathrm{Al}^{27}_{13}$) and requires $N_{\mathrm{POT}}^{\mathrm{Mu2e}}=\sn{3.6}{20}$ to achieve an single-event sensitivity of $\sn{2}{-17}$.  As mentioned in the previous section, we assume that the Mu2e-II reconstruction and selection efficiencies are unchanged relative to Mu2e so that the last term is equal to unity. The remaining quantities are taken from Tables~\ref{tab:nucleistuff} -- \ref{tab:1gev}.

We then take the resulting number of POT and calculate the implied beam power, and the number of muon and pion stops per kW, assuming the same run time for Mu2e-II as Mu2e, namely a three-year run with $\sn{2}{7}$ seconds of run time per year with proton pulse spacing of 1695~ns.  
%
%
The results of these calculations are given in Table~\ref{tab:npot}.

\begin{table}[tb]
  \centering
  \begin{tabular}{llcccc} \hline\hline
  & &\hspace*{0.15in} & 3~\gev\ &\hspace*{0.15in} & 1~\gev\ \\ \hline
  \multirow{4}{*}{Aluminum (3y run)\hspace*{0.10in}} 
    & $N_{\mathrm{POT}}$ & & \sn{8.7}{21} & & \sn{4.0}{22} \\
    & beam power (kW)    & & 72           & & 112 \\
    & $\mu$ stops / kW   & & \sn{8.0}{16} & & \sn{5.2}{16} \\   
    & $\pi$ stops / kW   & & \sn{3.5}{13} & & \sn{2.3}{13} \\ \hline
  \multirow{4}{*}{Titanium (3y run)\hspace*{0.10in}} 
    & $N_{\mathrm{POT}}$ & & \sn{1.1}{22} & & \sn{5.3}{22} \\
    & beam power (kW)    & & 94           & & 147 \\
    & $\mu$ stops / kW   & & \sn{8.0}{16} & & \sn{5.2}{16} \\   
    & $\pi$ stops / kW   & & \sn{3.5}{13} & & \sn{2.3}{13} \\ \hline\hline
  \end{tabular}
  \caption{Calculated number of POT necessary for Mu2e-II to achieve a 
    $\times 10$ smaller single-event-sensitivity than Mu2e and the
    corresponding beam power and muon and pion stopping rates.  The beam
    power is calculated assuming the same run time (three-year run with 
    $\sn{2}{7}$ s/year) and proton pulse spacing (1695 ns) as Mu2e.
  }
  \label{tab:npot}
\end{table}

\subsection{Estimate of instantaneous rates}
\label{sec:rates}

%
%

To aid in understanding what components of the currently planned Mu2e will require upgrading for a Mu2e-II experiment the beam-related instantaneous rates at the detector elements were estimated for the Project-X scenarios.
Although the beam power increases by a factor of 10 or more, the instantaneous rates only increase by a factor of 3-5 owing to the increased duty factor expected from Project-X.  This has important consequences for the viability of reusing much of the currently planned Mu2e detector apparatus for a next generation Mu2e-II.  For example, the current Mu2e tracker is being designed to handle instantaneous rates higher than what is currently estimated.  Simulation studies in which the instantaneous rates are increased by factors of two or four have been performed.  These studies indicate that at four times the nominal rates the tracker reconstruction efficiency only falls by about 5\% while maintaining the same momentum resolution.  These studies indicate that the Mu2e tracker would be able to handle the Project-X rates. 


\subsection{Anti-proton-induced backgrounds}
\label{sec:pbar}

There are no anti-proton-induced backgrounds for the Project-X scenarios since both 1~\gev\ and 3~\gev\ kinetic energy protons are below the anti-proton production threshold.
%

\subsection{Cosmic-ray-induced background}
\label{sec:cray}

The cosmic ray background is independent of $N_{\mathrm{POT}}$ and beam power and depends only on the live time of the experiment. Since we assume that Mu2e-II will have the same run time (three-year run with $\sn{2}{7}$ s/year) and proton pulse spacing (1695~ns) as Mu2e, the only difference is in the duty factor and the live-gate fraction. We scale the current Mu2e estimate for the cosmic-ray-induced backgrounds to account for these differences.
%
We assume that the veto efficiency for Mu2e-II is unchanged relative to Mu2e (99.99\% ). For a Project-X (PX) driven Mu2e-II experiment, the resulting cosmic-ray-induced background is $B_{\mathrm{CR}}^{\mathrm{PX}}=0.16$ events. In order to keep 
$B_{\mathrm{CR}}^{\mathrm{PX}} = B_{\mathrm{CR}}^{\mathrm{Mu2e}}$
the veto inefficiency would have to be reduced by a factor of three
to $\sn{3}{-5}$ so that the veto efficiency increases to 99.997\%.

\subsection{Radiative pion capture background}
\label{sec:rpc}

There are two components to the radiative pion capture (RPC) background: i) from pions produced from in-time protons interacting with the production target and ii) from pions produced from out-of-time protons interacting with the production target.  These are calculated separately.  

The probability that the photon from an RPC generates a background event depends on the fraction of pion captures that produce a photon, the probability that the photon will interact in the stopping target to produce a high energy electron, and the reconstruction and selection efficiency for those electrons.  We assume that the product of these factors, $\alpha_{\mathrm{RPC}} = \sn{1}{-6}$~\cite{Mu2eCDR}, which includes the contribution from the internal conversion process, in Mu2e-II is unchanged relative to Mu2e.  

The contribution to the RPC background from in-time protons is estimated using
\begin{equation}
  B_{\mathrm{RPC}}^{\mathrm{PX}}(\mathrm{in\: time}) = 
    S^{\mathrm{PX}}_{\pi} \cdot N_{\mathrm{POT}}^{\mathrm{PX}} \cdot
    F^{\mathrm{PX}}_{\pi} \cdot \alpha_{\mathrm{RPC}},
\end{equation}
where $S_\pi$ is the stopped-pion yield and $F_\pi$ is the fraction of pion captures in the live gate, both taken from Tables~\ref{tab:3gev} and~\ref{tab:1gev}, and $N_{\mathrm{POT}}$ is taken from Table~\ref{tab:npot}.

The contribution to the RPC background from out-of-time protons is estimated using
\begin{equation}
  B_{\mathrm{RPC}}^{\mathrm{PX}}(\mathrm{out\: of\: time}) = 
    S^{\mathrm{PX}}_{\pi} \cdot N_{\mathrm{POT}}^{\mathrm{PX}} \cdot
    E^{\mathrm{PX}} \cdot L^{\mathrm{PX}} \cdot \alpha_{\mathrm{RPC}},
\end{equation}
where $E$ is the extinction and $L$ is the live gate fraction from 
Table~\ref{tab:Mu2eBeamParameters}.  Here, as in~\cite{Mu2eCDR} we have assumed that the out-of-time protons are uniformly distributed in time.
The total RPC background estimate for the various Project-X (PX) scenarios is given in Table~\ref{tab:rpcBgd}.  Note that the RPC background is kept under control for Mu2e-II owing to the narrower proton pulse widths at Project-X and owing to the significantly improved extinction provided by Project-X beams.

\begin{table}[tb]
  \centering
  \begin{tabular}{llcccc} \hline\hline
  & &\hspace*{0.15in} & 3~\gev\ &\hspace*{0.15in} & 1~\gev\ \\ \hline
  \multirow{3}{*}{Aluminum\hspace*{0.10in}}
    & RPC from in-time p & & \sn{2.9}{-2} & & \sn{3.6}{-2} \\
    & RPC from late p    & & \sn{1.4}{-3} & & \sn{1.5}{-3} \\
    & RPC total          & & \sn{3.0}{-2} & & \sn{3.8}{-2} \\ \hline
  \multirow{3}{*}{Titanium\hspace*{0.10in}}
    & RPC from in-time p & & \sn{3.7}{-2} & & \sn{4.8}{-2} \\
    & RPC from late p    & & \sn{1.9}{-3} & & \sn{2.0}{-3} \\
    & RPC total          & & \sn{3.9}{-2} & & \sn{5.0}{-2} \\ \hline\hline
  \end{tabular}
  \caption{Estimates of the RPC background for various Project-X 
    scenarios.
  }
  \label{tab:rpcBgd}
\end{table}

\subsection{Other late arriving backgrounds}
\label{sec:late}

In addition to the RPC background discussed in the previous sub-section, out-of-time protons which arrive at the production target late can also give rise the $\mu$ decay-in-flight, $\pi$ decay-in-flight, and beam electron backgrounds.  For Mu2e-II these are estimated scaling from the Mu2e estimates and accounting for differences in the total number of protons on target, the yield of stopped-pions per proton, the extinction ratio, and the live-gate fraction between Mu2e and the Mu2e-II scenarios considered here.
The resulting estimated backgrounds for Mu2e-II from late arriving protons are given in Table~\ref{tab:LateBgd}.  
These backgrounds are kept under control for Mu2e-II since the increase in 
the total number of stopped pions is more than offset by the decrease in the extinction ratio relative to those same quantities for Mu2e.

\begin{table}[tb]
  \centering
  \begin{tabular}{llcccc} \hline\hline
  & &\hspace{0.15in} & 3~\gev\ &\hspace*{0.15in} & 1~\gev\ \\ \hline
  \multirow{3}{*}{Aluminum\hspace*{0.10in}}
    & $\mu$ decay in flight & & \sn{3.2}{-4} & & \sn{3.2}{-4} \\
    & $\pi$ decay in flight & & \sn{3.0}{-4} & & \sn{3.1}{-4} \\
    & beam electrons        & & \sn{4.4}{-4} & & \sn{4.6}{-4} \\ \hline
  \multirow{3}{*}{Titanium\hspace*{0.10in}}
    & $\mu$ decay in flight & & \sn{4.1}{-4} & & \sn{4.2}{-4} \\
    & $\pi$ decay in flight & & \sn{3.9}{-4} & & \sn{4.1}{-4} \\
    & beam electrons        & & \sn{5.8}{-4} & & \sn{6.0}{-4} \\ \hline\hline
  \end{tabular}
  \caption{Estimates of backgrounds from late arriving protons for various 
    Project-X scenarios. The RPC background from late arriving protons is
    discussed in Sec.~\ref{sec:rpc}.
  }
  \label{tab:LateBgd}
\end{table}

\subsection{Radiative muon capture background}
\label{sec:rmc}

The background from radiative muon captures (RMC) is estimated using the measured fraction of muon captures that result in a photon with energy larger than 53~MeV, $f_{\mathrm{RMC}}=\sn{1.43}{-5}$~\cite{rmcFraction}, and assuming that the probability for an RMC event to survive the reconstruction and selection requirements for Mu2e-II is unchanged relative to Mu2e, $\alpha_{\mathrm{RMC}} = \sn{7.8}{-19}$~\cite{Mu2eCDR}. 
We account for differences in the total number of protons on target, the yield of stopped muons, and the live gate fraction between Mu2e and Mu2e-II.  We also account for differences in the capture fraction when considering various stopping-target nuclei.
The resulting estimates are given in Table~\ref{tab:rmcBgd}.  Note that this is an intrinsic background that scales linearly with the number of stopped muons, in the same way the sensitivity does.  It is a negligible background for Mu2e and remains so for Mu2e-II. 

\begin{table}[tb]
  \centering
  \begin{tabular}{llcccc} \hline\hline
  & &\hspace*{0.15in} & 3~\gev\ &\hspace*{0.15in} & 1~\gev\ \\ \hline
  \multirow{3}{*}{Aluminum\hspace*{0.10in}}
    & $\mu$ captures in live gate & & \sn{1.8}{18} & & \sn{1.8}{18} \\
    & RMC in live gate            & & \sn{2.5}{13} & & \sn{2.5}{13} \\
    & RMC background              & & \sn{2.0}{-5} & & \sn{2.0}{-5} \\ \hline
  \multirow{3}{*}{Titanium\hspace*{0.10in}}
    & $\mu$ captures in live gate & & \sn{1.7}{18} & & \sn{1.7}{18} \\
    & RMC in live gate            & & \sn{2.5}{13} & & \sn{2.5}{13} \\
    & RMC background              & & \sn{1.9}{-5} & & \sn{1.9}{-5} \\ \hline\hline
  \end{tabular}
  \caption{Estimates of the radiative muon capture background for various 
    Project-X scenarios.
  }
  \label{tab:rmcBgd}
\end{table}

\subsection{Decay-in-Orbit background}
\label{sec:dio}

The $\mu$ decay-in-orbit background is estimated using fully simulated events including occupancy effects arising from DIO decays as well as from protons, neutrons, and photons from the muon captures in the stopping target.
Before producing any new simulations we first verified that we could reproduce the DIO background and corresponding single-event-sensitivity reported in the Mu2e CDR~\cite{Mu2eCDR} using the default Mu2e geometry and expected occupancies.  Then we simulated two separate geometries including the effects of the additional occupancy expected for the Mu2e-II scenarios discussed here.  The first geometry matches the current Mu2e design that features a straw tracker employing 15~$\mu m$ thick straw walls, while the second geometry matches the current Mu2e design geometry but assumes 8~$\mu m$ thick straw walls.  This second geometry is motivated by the fact that the Mu2e momentum resolution is scattering dominated with scattering in the straw walls a significant contribution.  It is also an advance in technical capabilities that seems plausible to accomplish with a few years of R \& D.  

For each geometry we simulated an aluminum stopping target and a titanium stopping target.  The shapes of the aluminum and titanium DIO spectra are taken from the latest calculations provided in reference~\cite{czarnecki}.  The resulting estimates of the DIO background are given in Table~\ref{tab:dioBgd}.  For these estimates the single-event-sensitivity is held fixed so that it's ten times smaller than the currently planned Mu2e.  While a rigorous optimization was not performed, the width of the momentum window used to select signal events was varied separately in each case and the numbers reported correspond to the window that roughly minimized the DIO background yield while not sacrificing too much signal efficiency.  The signal efficiencies match that reported in reference~\cite{Mu2eCDR} within about $15\%$ (relative).  

Note that the muon DIO process is an intrinsic background that scales linearly with the number of stopped muons in the same way as the sensitivity.  Thus, it is expected that for the same momentum resolution, a Mu2e-II experiment with an aluminum stopping target and a $\times 10$ better sensitivity than Mu2e will have a DIO background $\times 10$ larger than Mu2e.  This is reflected in Table~\ref{tab:dioBgd}.  The main difference between the aluminum and titanium DIO spectra is that near the kinematic endpoint, which is the region of interest, the titanium spectrum is falling more steeply than for aluminum.  This makes it more difficult to separate the DIO background from the signal in titanium.  This, too, is reflected in Table~\ref{tab:dioBgd}.
The only mitigations are to make more stringent momentum requirements or to redesign the tracker and the upstream material (e.g. stopping target, internal proton and neutron absorbers, etc.) to improve the momentum resolution.  Studies of the DIO background performed for the Mu2e experiment show that
it's unlikely that more stringent selection requirements alone could mitigate a ten-fold increase in the DIO background without a significant reduction in signal sensitivity.  

\begin{table}[t]
  \centering
  \begin{tabular}{lccc} \hline\hline
  & &\hspace*{0.15in} & DIO Yield \\ \hline
  \multirow{2}{*}{Aluminum\hspace*{0.10in}}
    & 15~$\mu m$ straws & & 2.14  \\
    & 8~$\mu m$ straws  & & 0.26  \\ \hline
  \multirow{2}{*}{Titanium\hspace*{0.10in}}
    & 15~$\mu m$ straws & & 2.25  \\
    & 8~$\mu m$ straws  & & 1.19  \\ \hline
  \end{tabular}
  \caption{Estimates of $\mu$ decay-in-orbit background for Project-X 
    scenarios using the current Mu2e tracker (15~$\mu m$) and assuming an 
    upgraded tracker with straw walls half as thick (8~$\mu m$).  The 
    yields reported here are the same (within $10\%$) for the 1~GeV and 
    3~GeV beam energies.
  }
  \label{tab:dioBgd}
\end{table}

\subsection{Background Uncertainties}
\label{sec:BgdUncertainties}

The total uncertainty on the mean expected background yield for Mu2e is about 20\%.  There are several contributing sources of uncertainty affecting these estimated background yields.  Most arise from uncertainties in production cross-section, in modeling, or in acceptance effects and are expected to contribute at a comparable level for Mu2e-II. A potential problem might arise from the precision with which the absolute momentum scale of the spectrometer is known.  This momentum-scale uncertainty affects the background yield estimates and is most pronounced for the DIO background since the DIO electron energy spectrum changes rapidly for energies near $105$~MeV.  Mu2e simulations estimate that a $\pm 0.08$ ($\pm 0.04$)~MeV momentum-scale uncertainty gives a $\pm 0.1$ ($\pm 0.05$)~event uncertainty on the DIO background yield, which corresponds to a 50\% (25\%) relative uncertainty.  The Mu2e experiment has as its goal to determine the momentum scale with a precision of $< 0.04$~MeV.  For the increased sensitivity of a Mu2e-II experiment an improved momentum-scale calibration may be required.

\FloatBarrier
\section{Discussion of facility upgrades}
\label{sec:upgrades}

We briefly discuss which elements of the currently planned Mu2e facility
it will be necessary to upgrade for the Mu2e-II scenarios discussed above.  We briefly discuss the target, the solenoids, the tracker, the calorimeter, the cosmic-ray veto, and the neutron/photon shielding systems.  Although they are not discussed explicitly here, it's clear that upgrades to the DAQ/Trigger would likely be required and the viability of the stopping-target monitor would likely have to be revisited.  We'll also note here that the radiation shielding in the proton beamline upstream of the production target would certainly have to be upgraded to handle the higher rates discussed in Sec.~\ref{sec:rates}.  For the lower beam energies considered here, it would also be necessary to reconfigure the proton beamline in order to center the proton beam on the production target.

\subsection{Target considerations}
\label{sec:targethall}

To accommodate beam power in the 80kW-110kW range estimated for Mu2e-II several aspects of the Target and Target Hall would have to be upgraded.  The proton beam dump would need improved cooling and the production target would need to be redesigned.  A new production target design would likely require modifications to the remote target handling system.  Depending on the proton beam energy and the configuration of the production solenoid, it may be necessary to redesign some portions of the Mu2e beam line just upstream of the production solenoid. Since the extinction in the Project-X scenarios is a factor 100 smaller than for Mu2e, the Extinction Monitor system would have to be upgraded to increased acceptance/sensitivity.

\subsection{Solenoid considerations}
\label{sec:solenoids}

It is expected that even at the increased rates and beam power discussed in Sec.~\ref{sec:rates}, the transport solenoid and detector solenoid should be able to operate reliably for a Mu2e-II run. 

\begin{figure}[hbt]
  \centering
  \includegraphics[width=0.39\textwidth]{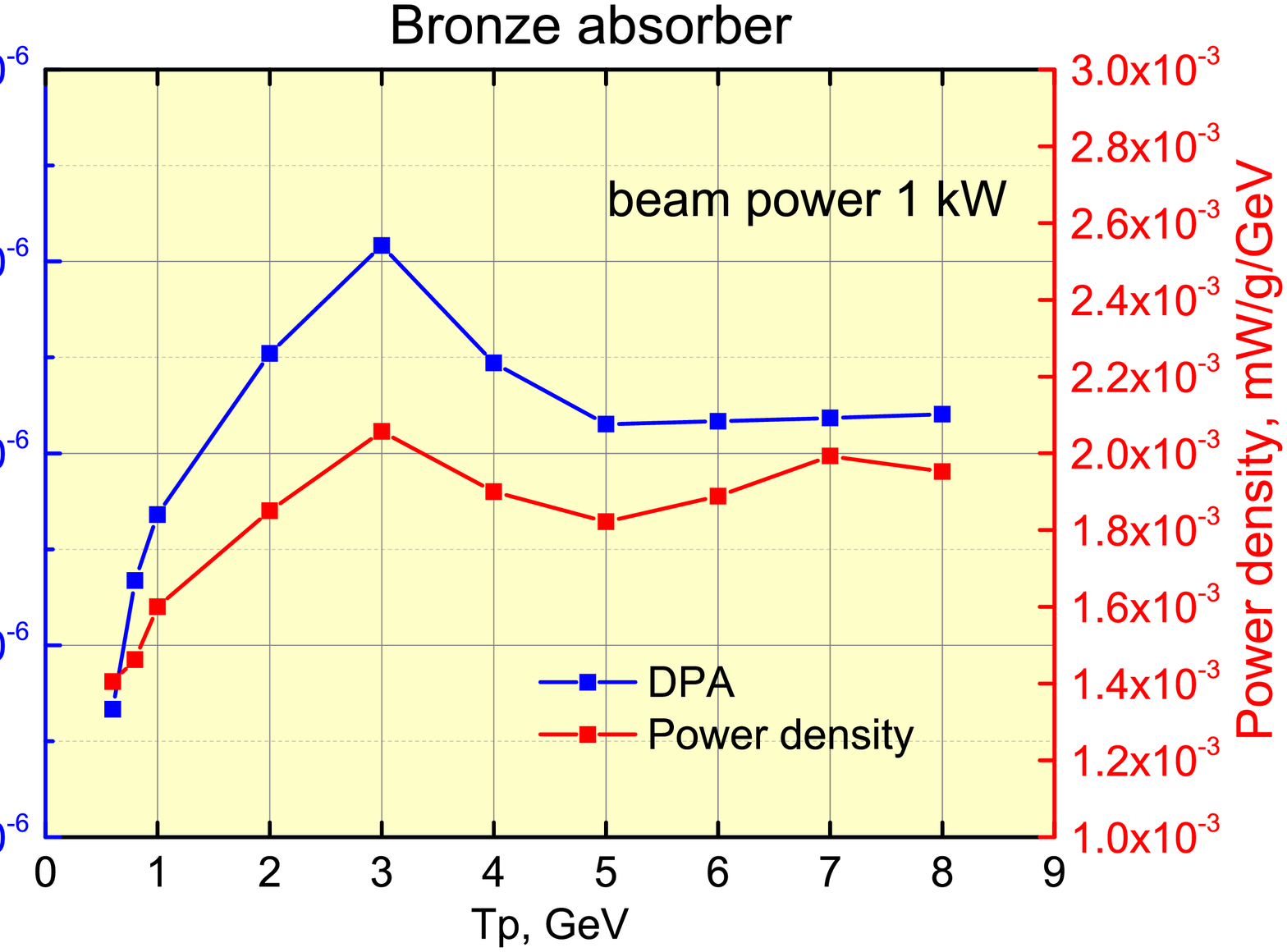}
  \includegraphics[width=0.47\textwidth]{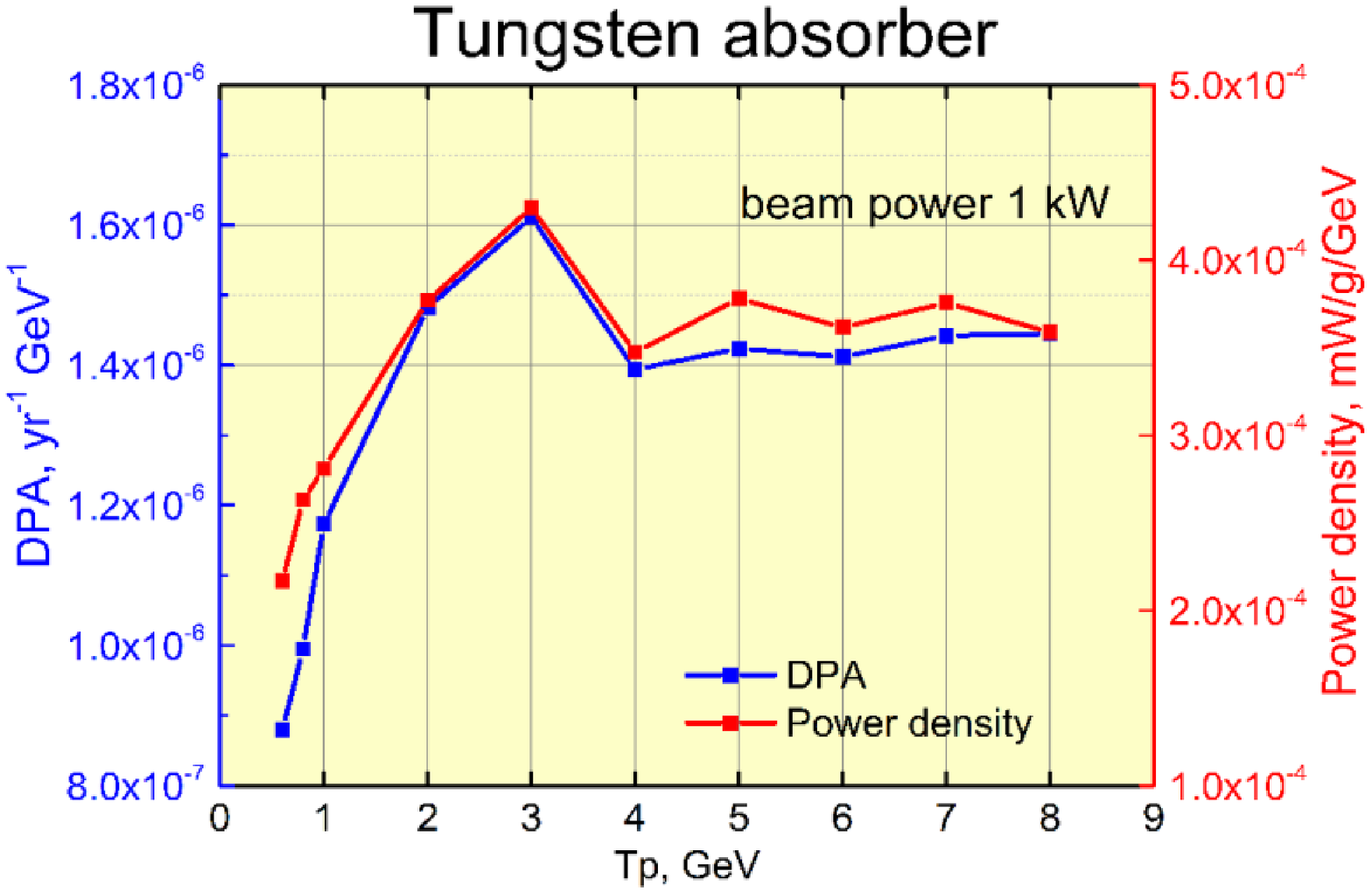}
  \caption{The peak power and peak dpa depositions for a 1 kW proton beam are shown as a function of the proton beam energy.  The plot on the left assume an all bronze HRS, as currently planned for Mu2e.  The plot on the right assumes an all tungsten HRS, as might be used for Mu2e-II.  The curves are estimates made using MARS~\cite{MARS} including a detailed model of the Mu2e solenoids, magnetic field, and beamline elements.
  }
  \label{fig:dpa}
\end{figure}

For the production solenoid, the limiting factors are the peak power and radiation damage (expressed in terms of displacements-per-atom or dpa) it would incur at the increased beam power.  To mitigate these effects a heat-and-radiation shield (HRS), made of bronze, will be installed along the inner walls of the production solenoid.  The magnet is designed to withstand a peak power deposition of about $30$~$\mu$W/g and a peak dpa of about $5\times 10^{-5}$~dpa/yr.  As demonstrated in Fig.~\ref{fig:dpa}, this bronze HRS would be inadequate for beam powers around 100 kW as required for the Mu2e-II scenarios considered here.  The performance of a tungsten HRS of the same geometry is given in Fig.~\ref{fig:dpa} and shows that such an HRS would adequately mitigate the peak-power deposition, it would allow a peak-dpa that is about a factor of three larger than the Mu2e requirements.  The main effect of an increased dpa is a faster degradation of the conductor residual resistivity ratio (RRR), which affects the solenoid's ability to operate reliably.  The current Mu2e production solenoid is designed to reliably operate with RRR$>100$.  Falling below this number does not immediately impact the solenoid performance, although once the RRR is low enough frequent quenches can result.  This is clearly a situation to be avoided.  The RRR can recover via a room temperature anneal.  The current Mu2e requirements for peak-dpa are set so that the experiment is unlikely to have to perform a room temperature anneal in a time period shorter than 12 months.  A significantly larger peak-dpa would require frequent room temperature annealing and compromise the experiment up time.  Consequently, for the Mu2e-II scenarios considered here, a re-optimization of the HRS geometry would have to be pursued.  If the peak-dpa could not be further reduced, it may be necessary to replace the production solenoid with a more radiation tolerant design.

In addition to the HRS, several solenoid sub-systems would have to be revisited and possibly upgraded.  In particular the cooling system and cryogenics plant would be reevaluated in light of the increased heat load.  If the PS is replaced, the related power supplies, bus work, feedbox, and transfer lines would be reevaluated.

\subsection{Tracker considerations}
\label{sec:tracker}

The Mu2e tracker provides the primary momentum measurement for charged particles originating in the stopping target. The tracker must accurately and efficiently identify and measure 105 MeV/c electrons while rejecting backgrounds. The Mu2e tracker is a low mass array of straw drift tubes aligned transverse to the axis of the detector solenoid. The basic detector element is a straw made of a 25 $\mu m$ sense wire inside a 5 $mm$ diameter tube made of 15 $\mu m$ thick metalized Mylar. The detector will have about 20k straws evenly distributed among 18 measurement stations along a length of about 3 meters.
When considering the expected performance of the Mu2e tracker in the context of Mu2e-II, there are two major issues: the tracker performance, and the tracker operability.

To understand the tracker performance at the increased rates discussed in Sec.~\ref{sec:rates} a simulation was performed in which the tracker occupancy arising from beam-related background processes is increased by a factor of four.  A signal sample of electrons from $\mu^- N\rightarrow e^- N$ conversion and a sample of electrons from muon DIO were generated with the higher occupancy and reconstructed using the standard Mu2e algorithms.  Using the selection criteria that give the sensitivity discussed in Sec.~\ref{sec:introduction} for Mu2e, we find in the high occupancy samples that the reconstruction efficiency drops by only 5\% while the momentum resolution -- and therefore the DIO background -- are unchanged.  It should be noted that some of the lost efficiency may be recoverable with improvements to the reconstruction and track-fitting algorithms.  We conclude that the performance of the current Mu2e straw tracker design is not significantly affected by the rates discussed here for a future Mu2e-II experiment.  

Concerning the tracker operability, the most important challenges to consider are related to the beam flash.  Space-charge effects can arise
with sufficiently high occupancies because charge accumulates near the straw wall -- due to the relatively slow drift of ions -- and screens the cathode to produce a drastic decrease in gain. Based on the expected rates discussed here for Mu2e-II we estimate that only $\approx 1\%$ of straws will suffer from this, concentrated on the inner straw layer.  High rates from the beam flash can also cause high enough current draws that the high voltage sags and gain is reduced.  At the Mu2e-II rates considered here the estimated high-voltage sag is negligible ($\approx$ 200mV) and can be easily mitigated by running the high voltage slightly higher than nominal. Finally, high currents and a large total dose may lead to aging effects in the straws. Although by extrapolating from current aging-test data we do not expect significant aging, a definitive measurement corresponding to the Mu2e-II intensities discussed will require an additional several months of exposure. There are several ways to mitigate all these issues. The most obvious, although perhaps difficult to implement, is to reduce the high voltage during the beam flash. Thinner straw walls will also significantly reduce all the effects mentioned above with the added benefit of improving the momentum resolution. Finally, the gain can be intentionally reduced at the center of the straws on the inner layer, which are the most affected, by plating a given length of the wire.
None of the problems listed above are insurmountable and although a strong R\&D program will be required to really demonstrate the proposed mitigations, these preliminary estimates suggest that the current Mu2e tracker design would be reliably operable in the rate environment discussed here for a future Mu2e-II experiment.

The momentum resolution of Mu2e is dominated by material effects; coulomb scattering in the tracker straw walls and energy straggling in the stopping target and proton absorber.  If, as the studies of the DIO background discussed in Sec.~\ref{sec:dio} suggest and depending on the outcome of the Mu2e experiment, an improved momentum resolution is required for Mu2e-II, then the tracker would have to be replaced.  The most straightforward replacement would employ the same design but use straws with thinner walls, perhaps as thin as 8~$\mu m$.    A simulation in which the straw walls were reduced to 8~$\mu m$ and the stopping-target and proton-absorber material halved yielded a momentum resolution about 50\% smaller than for Mu2e.  
A rigorous R\&D program would be required to demonstrate the viability of such thin-walled straws.

\subsection{Calorimeter considerations}
\label{sec:calorimeter}

The calorimeter plays an important role in mitigating potential backgrounds from muons, and from muons and electrons born downstream of the stopping target that travel upstream, get reflected back in the gradient field at the start of the detector solenoid, and are then reconstructed as they move back downstream as if having originated in the stopping target - most of these backgrounds are cosmic-ray-induced.  The calorimeter timing information, cluster shape, and energy measurement are all important in eliminating these backgrounds.  The efficacy of these veto criteria depends on the hit rates in the calorimeter from beam-related particles (e.g. photons and neutrons from the muon capture process, low energy electrons from the DIO process, etc).  Consequently the increased rates discussed in Sec.~\ref{sec:rates} will detrimentally affect the calorimeter's performance.  Simulation studies are necessary to quantify the degree to which the increased rates affect the calorimeter's performance.  If it turns out that the expected performance is not sufficient to meet the Mu2e-II physics requirements, then it may be necessary to upgrade to faster readout electronics or to a faster crystal (e.g. BaF$_2$).  Regardless of the outcome of the simulation studies, the photosensors will likely require replacement owing to radiation damage incurred during Mu2e running.  The performance of the LYSO crystals is not expected to be significantly affected by the radiation dose incurred during Mu2e running.

\subsection{Cosmic Ray Veto considerations}
\label{sec:crv}

A potentially dangerous background for Mu2e arises from cosmic-ray muons that interact or decay to produce an electron that mimics the signal.  For the Mu2e experiment, these backgrounds will be vetoed using an active shield composed of three layers of plastic scintillator, where the scintillation light is read out by wave-length-shifting fiber, and Silicon-based photo detectors (SiPMs).  Based on studies with the current Mu2e design, a cosmic ray veto (CRV) efficiency on the order of 99.99~\% is required to limit the background from cosmic rays to 0.05 events~\cite{Mu2eCDR}.  While, the cosmic ray background seems to be under control in the Mu2e experiment, it has been a limiting factor in past experiments~\cite{Ahmad:1988ur}, and must be carefully considered for any future experiment.
  
There are two major concerns with respect to the CRV when considering potential Mu2e upgrades.  The cosmic-ray-induced-background rate scales with the live time, unlike most backgrounds in the experiment, so for this concern high intensity and a low duty factor (i.e. low live time) are ideal.  A second concern is that neutron and gamma radiation induce accidental hits in the CRV and if the rate of accidentals is large enough overlaps can generate false vetoes and limit the experiment live time.  The neutron and gamma radiation also can damage the SiPMs and readout electronics.  For issues related to false vetoes caused by neutrons or gammas, low intensity and high duty factor are preferable.  The false-veto rate due to radiation can be reduced by requiring a higher threshold in each scintillator counter but this comes at the cost of reducing veto efficiency.  These tradeoffs between efficiency and dead time due to false vetoes must be kept in mind in the discussion below.

The initial Mu2e experimental run will provide a chance for careful studies of both of these issues in all regions of the the CRV.  The performance will be well measured in the first run using cosmic-ray data during periods with no beam and the rates from neutrons can be studied with beam.  Both can be used to tune simulation and improve the predictions for an experimental upgrade.  In addition, the veto algorithm can be further refined to improve performance beyond what is being used in current studies.    
 
Increasing the duty factor by a factor of three means that the cosmic ray muon background will increase by the same factor if the CRV and veto algorithm are held constant.  This may be acceptable given that the cosmic ray background will still be a sub-dominant background.  It is also possible that due to safety factors included in the CRV design that the CRV will be found to operate at a higher efficiency than the minimal requirements on the system and will be sufficient to suppress this background by an additional factor of three without adding significant dead time.  If additional efficiency is required, continued simulation efforts as well as experience with cosmic-ray data during the first run of Mu2e will identify the regions of the CRV coverage that require the highest efficiency, and upgrades to these regions such as thicker scintillator, additional layers, larger diameter fiber, or increasing the number of fibers can be considered to improve the rejection efficiency.        

Over most regions of the CRV, the increased rates considered here for Mu2e-II should not cause an issue with false-veto rates or radiation damage.  Radiation damage on the SiPMs should be manageable if the system is designed so that they survive the first run. That is, the worst case scenario would be that some SiPMs need to be replaced once per year in the highest-rate regions of the CRV.  A design where the light is routed to photo detectors located in a lower-radiation environment is currently being considered for the highest rate areas, and expanding this to more of the CRV coverage is a possibility.  On the relevant timescale, it's also likely that more radiation tolerant SiPMs will be available to replace what will be used for Mu2e. To reduce accidental rates, the highest-rate CRV coverage areas could be upgraded to finer segmentation.  Also, for regions requiring a reduced veto efficiency a higher threshold could be applied to reduce the accidental rate while still maintaining a high-enough veto efficiency.  Based on current simulation studies there is reason to believe that this strategy may be viable in some of the highest-rate regions of the CRV.
It is likely that there will be shielding optimizations that could help protect the CRV both from radiation damage and from unacceptably high false-veto rates.  

For a further intensity upgrade, say by a factor of 10, the scintillator-based CRV design may not be a viable option due to both high neutron incidental rates and radiation damage to the photo detectors and electronics, especially in the region of the transport solenoid. A more radiation-hard solid-state photo detector could be considered (GaAs for example).  Cathode strip chambers are an option that are more neutron blind, however dead-time rates due to gamma radiation must be well understood.  For this high-intensity scenario, it is clear that a significant upgrade to the shielding would be necessary in addition to other upgrades to the CRV technology.    

In summary, the current CRV design should be adaptable to the case with a factor of 3-4 increase in intensity necessary for Mu2e-II with only minor upgrades required and it is likely that these are only needed in the most intense radiation regions of the veto system.  A factor of 10 increase in intensity is more difficult troublesome, and improved shielding as well as another detector technology may be required in the highest radiation regions.  Experience from the first run of the Mu2e experiment will be crucial in quantifying limitations of the system for future intensity upgrades.

\subsection{Neutron and Photon Shielding considerations}
\label{sec:shielding}

Neutrons are copiously produced at several places along the Mu2e apparatus - at the tungsten production target, at the collimators located along the transport solenoid, at the aluminum stopping target, and at a muon-beam absorber at the upstream end of the detector solenoid.  These neutrons are sources of accidental rate in the tracker, calorimeter, and cosmic-ray veto system and cause radiation damage to various electronic components.  In addition, once the neutrons are thermalized, they are captured in the surrounding material.  The neutron capture processes often produce photons, which also contribute to the accidental rates in the detector systems.  To mitigate the effects of these neutrons and photons to acceptable levels, Mu2e employs several layers of various shielding materials both internal and external to the solenoid volumes.  For the higher Mu2e-II rates discussed in Sec.~\ref{sec:rates}, some of this shielding would have to be upgraded in order to sufficiently suppress the neutron and photon rates below acceptable levels.  

Possible shielding upgrades can be split between the two main
sources of neutrons/photons to the cosmic-veto counters: beam-induced backgrounds from the production-target region and muon-stopping-induced background from the aluminum stopping target and muon-beam absorber regions. Beam-related backgrounds from the production target region will be reduced per POT for the 1 and 3 GeV proton beam energies considered here. For these lower energies, studies of the neutron- and photon-induced rates at the CRV from beam interactions in the production solenoid and transport solenoid regions are being pursued. The muon-stopping-induced rates scale with the muon intensity and will affect the rates in the tracker and calorimeter in addition to the CRV rates.  In the detector solenoid region there is shielding inside the detector solenoid cryostat (the inner neutron shield), and external to the cryostat in the roughly 15 cm space between the cryostat and concrete CRV support walls.  The current inner neutron shield under consideration for Mu2e will be about 2200 kg of high-density-polyethylene shadowing the aluminum stopping target. This can be upgraded by changing material or adding mass.  Changing the inner neutron shield
to lithiated polyethylene reduces the photon rates in the tracker by about 33\% and the neutron rates in the CRV by about 25\%.  Simply increasing the mass of the inner neutron shield lowers the CRV neutron rate roughly proportional to the mass up to 3500 kg.  The external shielding can be upgraded by adding material in the space between the cryostat and the CRV support walls or by upgrading the support walls themselves.  Adding a water tank between the cryostat and support wall lowers the neutron rate by about two (with limited impact on the photon rate) while using either HDPE or stainless steel lowers the neutron and photon rates by about factors of four and two, respectively. Studies exploring the use of boronated and lithiated polyethylene in this region are being pursued. The initial Mu2e configuration may have no material in this region or high-density-polyethylene in the stopping target region only.  The initial Mu2e design uses 18 inches of normal concrete for the CRV support walls.  Changing this to "heavy" concrete (by adding either barium or iron) can reduce the CRV
neutron rate by about 2, while doubling the concrete thickness lowers
the neutron rate by about 10, depending on the neutron energy. These
changes to the concrete may only be needed near the aluminum stopping target and muon-beam absorber regions.  Changes in the thickness would necessitate changes to the CRV geometry in the affected regions.


\FloatBarrier
\section{Summary and Conclusion}
\label{sec:summary}

 We investigated the feasibility of a next generation Mu2e experiment (Mu2e-II) that uses as much of the currently planned facility as possible and Project-X beams to achieve a sensitivity that's about a factor of ten better than Mu2e.  We use a \texttt{G4Beamline} simulation to estimate the backgrounds assuming the currently planned Mu2e apparatus is viable with comparable performance at the Project-X scenarios explored.  The number of protons-on-target for each scenario is chosen to achieve a single-event-sensitivity of $\sn{2}{-18}$.  The run time and proton pulse spacing for Mu2e-II is assumed to be the same as Mu2e.  The resulting background estimates are given in Table~\ref{tab:PXBgd} along with the estimate for the currently planned Mu2e~\cite{DIOComment}.

Based on these studies we conclude that a Mu2e-II experiment that reuses a large fraction of the currently planned Mu2e apparatus and provides a $\times 10$ improved sensitivity is feasible at Project-X with 100-150 kW of 1 or 3~\gev\ proton beams.  Aside from the muon decay-in-orbit background, which requires improved momentum resolution to mitigate, the remaining backgrounds are kept under control due to important features of Project-X.  
The narrower proton pulse width and expected excellent intrinsic extinction are both important to mitigating the radiative-pion-capture background.  The excellent intrinsic extinction is also important in mitigating backgrounds from late arriving protons such as $\mu$ and $\pi$ decay-in-flight and beam-electron backgrounds.  A beam energy below the anti-proton production threshold eliminates the anti-proton induced backgrounds.  The high duty factor expected for Project-X is important since it enables a ten-fold improvement in sensitivity over a reasonable timescale while necessitating only a modest increase (a factor of 3-5, depending on scenario) in instantaneous rates at the detectors.   Because the instantaneous rates increase only modestly, we believe that Mu2e-II could reuse the currently planned Mu2e apparatus with only modest upgrades necessary.  To keep the DIO background below an event, it would be necessary to upgrade to a lower mass tracker.  Our simulations halved the thickness of the straw walls from 15~$\mu m$ to 8~$\mu m$, resulting is an improved momentum resolution and a reduction in the DIO background.

\begin{table}[t]
  \centering
  \begin{tabular}{llcrcrcr} \hline\hline
    & & \hspace*{0.15in} & \multicolumn{1}{c}{Mu2e} &\hspace{0.15in} &\multicolumn{3}{c}{Mu2e-II} \\
    & & & \multicolumn{1}{c}{8~\gev } & &\multicolumn{3}{c}{1 or 3~\gev } \\
    & & & \multicolumn{1}{c}{Al.} & & \multicolumn{1}{c}{Al.}  &\hspace*{0.1in} & \multicolumn{1}{c}{Ti.} \\ \hline
    Category      & Source                     
    & &\multicolumn{5}{c}{Events} \\ \hline
    \multirow{2}{*}{Intrinsic} 
                  & $\mu$ decay in orbit       
    & & 0.22 & & 0.26 & & $1.19$  \\ 
                  & radiative $\mu$ capture    
    & & $<0.01$ & & $<0.01$ & & $<0.01$  \\ \hline
    \multirow{4}{*}{Late Arriving}
                  & radiative $\pi$ capture    
    & & 0.03 & & 0.04 & & 0.05  \\
                  & beam electrons             
    & & $<0.01$ & & $<0.01$ & & $<0.01$  \\
                  & $\mu$ decay in flight      
    & & 0.01 & & $<0.01$ & & $<0.01$   \\
                  & $\pi$ decay in flight      
    & & $<0.01$ & & $<0.01$ & & $<0.01$  \\ \hline
    \multirow{3}{*}{Miscellaneous}
                  & anti-proton induced        
    & & 0.10 & & -- & & --  \\
                  & cosmic-ray induced         
    & & 0.05 & & 0.16 & & 0.16   \\
                  & pat. recognition errors    
    & & $<0.01$ & & $<0.01$ & & $<0.01$  \\ \hline
    Total Background &                         
    & & 0.41 & & 0.46 & & 1.40  \\ \hline\hline
  \end{tabular}
  \caption{A summary of the current Mu2e background estimate and 
    estimates of how the backgrounds would scale for a next generation
    Mu2e experiment, Mu2e-II, that employs Project-X beams at 1 or 
    3~\gev\ and an aluminum or titanium stopping target. For a given 
    stopping target, the difference in background yields between a 
    1~\gev\ or a 3~\gev\ proton beam is about 10\%.  The total 
    uncertainty on the total Mu2e background is estimated to be 
    about 20\%. Note that the DIO estimates for the Mu2e-II case
    assume the tracker has been upgraded to use thinner walled straws.
    If the tracker is not upgraded, the DIO estimates would increase as
    given in Table~\ref{tab:dioBgd} while the remaining backgrounds would
    not change significantly. 
  }
  \label{tab:PXBgd}
\end{table}


\FloatBarrier



\begin{thebibliography}{99}
\bibitem{MoveLiveGate} We investigated the possibility of also beginning the live gate at an earlier time, but this resulted in an unacceptable increase in background events from radiative pion captures.
\bibitem{DIOComment}
Note that the DIO estimate for the titanium stopping target is an extrapolation from a detailed simulation of the aluminum DIO spectrum scaled to obtain the correct endpoint for titanium.  Work is underway to include the titanium DIO spectrum in the simulation software and repeat this estimate.









\relax
\end{thebibliography}
\end{document}